\PassOptionsToPackage{unicode}{hyperref}
\PassOptionsToPackage{hyphens}{url}
\PassOptionsToPackage{dvipsnames,svgnames,x11names}{xcolor}
\documentclass[
  12pt]{article}

\usepackage{amsmath,amssymb,comment,rotating}
\usepackage{iftex}
\ifPDFTeX
  \usepackage[T1]{fontenc}
  \usepackage[utf8]{inputenc}
  \usepackage{textcomp} 
\else 
  \defaultfontfeatures{Scale=MatchLowercase}
  \defaultfontfeatures[\rmfamily]{Ligatures=TeX,Scale=1}
\fi
\usepackage{lmodern} 
\ifPDFTeX\else  
\fi
\IfFileExists{upquote.sty}{\usepackage{upquote}}{}
\IfFileExists{microtype.sty}{
  \usepackage[]{microtype}
  \UseMicrotypeSet[protrusion]{basicmath} 
}{}
\makeatletter
\@ifundefined{KOMAClassName}{
  \IfFileExists{parskip.sty}{%
    \usepackage{parskip}
  }{
    \setlength{\parindent}{0pt}
    \setlength{\parskip}{6pt plus 2pt minus 1pt}}
}{
  \KOMAoptions{parskip=half}}
\makeatother
\usepackage{xcolor}
\makeatletter
\ifx\paragraph\undefined\else
  \let\oldparagraph\paragraph
  \renewcommand{\paragraph}{
    \@ifstar
      \xxxParagraphStar
      \xxxParagraphNoStar
  }
  \newcommand{\xxxParagraphStar}[1]{\oldparagraph*{#1}\mbox{}}
  \newcommand{\xxxParagraphNoStar}[1]{\oldparagraph{#1}\mbox{}}
\fi
\ifx\subparagraph\undefined\else
  \let\oldsubparagraph\subparagraph
  \renewcommand{\subparagraph}{
    \@ifstar
      \xxxSubParagraphStar
      \xxxSubParagraphNoStar
  }
  \newcommand{\xxxSubParagraphStar}[1]{\oldsubparagraph*{#1}\mbox{}}
  \newcommand{\xxxSubParagraphNoStar}[1]{\oldsubparagraph{#1}\mbox{}}
\fi
\makeatother

\usepackage{longtable,booktabs,array}
\usepackage{calc} 
\usepackage{etoolbox}
\makeatletter
\patchcmd\longtable{\par}{\if@noskipsec\mbox{}\fi\par}{}{}
\makeatother
\IfFileExists{footnotehyper.sty}{\usepackage{footnotehyper}}{\usepackage{footnote}}
\makesavenoteenv{longtable}
\usepackage{graphicx}
\makeatletter
\def\maxwidth{\ifdim\Gin@nat@width>\linewidth\linewidth\else\Gin@nat@width\fi}
\def\maxheight{\ifdim\Gin@nat@height>\textheight\textheight\else\Gin@nat@height\fi}
\makeatother
\setkeys{Gin}{width=\maxwidth,height=\maxheight,keepaspectratio}
\makeatletter
\def\fps@figure{htbp}
\makeatother

\makeatletter
\@ifpackageloaded{caption}{}{\usepackage{caption}}
\AtBeginDocument{%
\ifdefined\contentsname
  \renewcommand*\contentsname{Table of contents}
\else
  \newcommand\contentsname{Table of contents}
\fi
\ifdefined\listfigurename
  \renewcommand*\listfigurename{List of Figures}
\else
  \newcommand\listfigurename{List of Figures}
\fi
\ifdefined\listtablename
  \renewcommand*\listtablename{List of Tables}
\else
  \newcommand\listtablename{List of Tables}
\fi
\ifdefined\figurename
  \renewcommand*\figurename{Figure}
\else
  \newcommand\figurename{Figure}
\fi
\ifdefined\tablename
  \renewcommand*\tablename{Table}
\else
  \newcommand\tablename{Table}
\fi
}
\@ifpackageloaded{float}{}{\usepackage{float}}
\floatstyle{ruled}
\@ifundefined{c@chapter}{\newfloat{codelisting}{h}{lop}}{\newfloat{codelisting}{h}{lop}[chapter]}
\floatname{codelisting}{Listing}

\makeatother
\makeatletter
\@ifpackageloaded{caption}{}{\usepackage{caption}}
\@ifpackageloaded{subcaption}{}{\usepackage{subcaption}}
\makeatother

\ifLuaTeX
  \usepackage{selnolig}  
\fi
\usepackage[]{natbib}
\usepackage{bookmark}
\usepackage{multirow}
\usepackage{bm}
\usepackage{setspace}
\IfFileExists{xurl.sty}{\usepackage{xurl}}{} 
\hypersetup{
  pdftitle={Title},
  pdfauthor={Author 1; Author 2},
  pdfkeywords={3 to 6 keywords, that do not appear in the title},
  colorlinks=true,
  linkcolor={blue},
  filecolor={Maroon},
  citecolor={Blue},
  urlcolor={Blue},
  pdfcreator={LaTeX via pandoc}}

\usepackage{algorithm}
\usepackage{algpseudocode}

\newcommand{\anon}{1}
\newcommand{\Yb}{\mathbf{Y}}
\newcommand{\Yi}{\bm{Y}}
\newcommand{\yb}{\mathbf{y}}

\newtheorem{lemma}{Lemma}

\begin{document}

\def\spacingset#1{\renewcommand{\baselinestretch}%
{#1}\small\normalsize} \spacingset{1}


\if1\anon
{
  \title{\bf Bayesian Graphical Models under Positivity Constraints: A Scalable Generalized Likelihood Approach}
   \author{Swarnali Raha \hspace{.2cm}\\
   Department of Statistics, University of Florida\\
    Partha Sarkar \\
   Department of Statistics, Florida State University\\
   Sirani M. Perera\\
   Mathematics Department, Embry-Riddle Aeronautical University\\
    and \\
    Kshitij Khare\\
   Department of Statistics, University of Florida}
  \maketitle
} \fi

\if0\anon
{
  \bigskip
  \bigskip
  \bigskip
  \begin{center}
    {\large \bf Bayesian Graphical Models under Positivity Constraints: A Scalable Generalized Likelihood Approach}
\end{center}
  \medskip
} \fi

\bigskip
\begin{abstract}
\noindent We develop a computationally scalable Bayesian framework for precision matrix estimation in Gaussian graphical models under total positivity constraints. To overcome the high computational cost of the Gaussian likelihood, we adopt a generalized Bayesian approach based on the D-trace loss, which eliminates the log-determinant term and enables efficient optimization while allowing relaxation of positive definiteness during sampling. Sparsity is induced via spike-and-slab priors, and the resulting generalized posterior is shown to be proper under mild conditions. Our primary contribution is a suite of efficient posterior sampling algorithms tailored to high-dimensional settings. Starting from a component-wise Gibbs sampler, we introduce a novel data augmentation scheme that induces conditional independence among precision matrix entries, enabling joint updates. By exploiting the Gram structure of the sample covariance matrix, we further develop a fast matrix-normal sampler that significantly reduces per-iteration complexity in high-dimensional settings. An interweaving strategy combines augmented and direct updates to improve mixing without sacrificing scalability. Experiments on synthetic and financial data demonstrate substantial computational gains over existing methods, while maintaining competitive estimation accuracy and improved recovery of structured dependencies. 
\end{abstract}

\noindent%
{\it Keywords: Gaussian graphical models; 
total positivity; generalized Bayesian inference; data augmentation; Gibbs sampling.} 
\vfill

\spacingset{1.8} 

\section{Introduction}\label{sec:1:intro}

Graphical models represent dependence among random variables through a graph, where vertices correspond to variables and edges encode conditional associations. Let $\Yi=(Y_1,\ldots,Y_p)^\top$ be a $p$-variate random vector with covariance matrix $\Sigma$ and precision matrix $\Omega=\Sigma^{-1}=[\omega_{ij}]_{i,j=1}^{p}$. Under a Gaussian model, $\omega_{ij}=0$ if and only if $Y_i$ and $Y_j$ are conditionally independent given all remaining variables. More generally, zeros in $\Omega$ correspond to vanishing partial correlations. Thus, graph estimation reduces to identifying the zero and nonzero off-diagonal entries of $\Omega$. 

The high-dimensional setting, where $p$ is comparable to or exceeds the sample size $n$, has motivated extensive work on sparse precision matrix estimation under the assumption of a sparse underlying graph~\citep{glasso2009,ravikumar2011l1,LamFan2009,friedman2010applications,Cai01062011,Ni2022GGMM,Behdin2023GraphL0,Shiratori2024DCGGM,Bhadra2022Telescoping}. 

A particularly important class of graphical models imposes total positivity constraints \citep{ying2023adaptiveestimationgraphicalmodels,lauritzen_total_positivity1,lauritzen_total_positivity2,slawski,wang2020learninghighdimensionalgaussiangraphical}. In the Gaussian setting, this is equivalent to restricting $\Omega$ to be an $M$-matrix, that is, a positive definite matrix with nonpositive off-diagonal entries. Such models imply nonnegative partial correlations and constitute the Gaussian specialization of the broader $MTP_2$ framework, which encodes strong positive dependence \citep{lauritzen_total_positivity1,lake_tenenbaum_2010,Agrawal,article}. These dependence structures arise naturally in applications including finance, psychometrics, cognition, and signal processing. 

Despite substantial progress on sparse graphical models, most work under total positivity constraints is frequentist. Bayesian methods, while offering uncertainty quantification, remain relatively underdeveloped. Existing Gaussian-likelihood-based samplers become difficult to scale in the presence of sign constraints, as enforcing the $M$-matrix restriction requires sampling from high-dimensional truncated distributions. This often leads to slow mixing and substantial computational overhead. Moreover, even in unconstrained settings, Bayesian samplers typically have per-iteration complexity $\mathcal O(p^4)$, with reductions to $\mathcal O(p^3)$ only under restricted regimes \citep{gao2026ordermagnitudetimecomplexity} (such as $n = o(\sqrt{p})$), making the constrained setting even more computationally challenging.  

To address these challenges, we develop a generalized Bayesian framework based on the D-trace loss, replacing the Gaussian likelihood with a computationally tractable surrogate. Such loss-based formulations avoid the log-determinant term and have been widely used in graphical modeling \citep{buhlman,Peng,rocha2008,khare2015convex,Cai01062011,4b5cd03a-e599-3ef7-a1a6-896148b7af15}. Combined with spike-and-slab priors \citep{george1993variable}, the resulting generalized posterior is shown to be proper and supports both sparsity learning and uncertainty quantification. 

We first develop a direct element-wise Gibbs sampler for posterior computation. This direct sampler is straightforward and has per-iteration complexity $\mathcal{O}(p^3)$ for arbitrary $n$ and $p$, which is lower than that of standard Gaussian-likelihood-based samplers. Nevertheless, this approach can still be computationally challenging in the high-dimensional regime $n \ll p$. Moreover, since the elements of the precision matrix are updated sequentially and are typically strongly dependent, the resulting MCMC updates may exhibit suboptimal results (see Section \ref{sec:4:methods}). 

To alleviate these computational bottlenecks, we introduce a data-augmentation scheme motivated by the anti-correlated Gaussian approach of \cite{zheng2023gibbs}. This framework was originally developed for multivariate regression, where augmentation reduces posterior dependence by inducing conditional independence. Generalizing this idea to the matrix-valued setting is nontrivial and requires a substantial reformulation; see Section~\ref{subsec:4.2:aug:gibbs}. The resulting augmentation yields conditional independence among the entries of the precision matrix given the augmented variables, leading to the augmented Gibbs sampler in Algorithm \ref{algo:2:auggibbs}. 

Our augmented Gibbs sampler enjoys several benefits. First, it improves computational scalability relative to the direct sampler, particularly in high-dimensional regimes. While Gaussian-likelihood-based (unconstrained) samplers typically scale as $\mathcal{O}(p^4)$, with recent improvements \citep{gao2026ordermagnitudetimecomplexity} achieving $\mathcal{O}(p^3)$ under $n=o(\sqrt p)$, our method has per-draw complexity $\mathcal{O}(min\{p^2 n, p^3\})$, which becomes $o(p^{2.5})$ in the same regime. Second, the D-trace loss, together with the relaxation of the positive-definiteness constraint and the induced conditional independence, allows the $M$-matrix constraint to be incorporated naturally.

Finally, motivated by \cite{Yu01012011}, we propose an interweaving Gibbs sampler that combines the efficiency of the augmented sampler with occasional updates from the direct sampler under the original parameterization. The augmented sampler enables efficient global exploration, while the direct sampler provides complementary local updates. Empirically, this yields improved stability and mixing with minimal additional cost (see Algorithm~\ref{algo:3:interweaving} and Section~\ref{subsec:4.3:interweaving}).

Simulation studies demonstrate that the proposed methods achieve competitive, and often superior, performance in graph recovery and precision matrix estimation while maintaining scalability in high-dimensional settings. In an analysis of S\&P 500 returns, the estimated networks recover sectoral structure more effectively than competing approaches, yielding higher modularity scores and clearer within-sector connectivity. 

The remainder of the paper is organized as follows. Section~\ref{sec:2:setup} introduces the notation, background on generalized likelihood approaches, and the D-trace loss. Section~\ref{sec:3:model} presents the proposed generalized Bayesian framework based on the D-trace loss and spike-and-slab priors, and establishes the propriety of the resulting (generalized) posterior. Section~\ref{sec:4:methods} develops the posterior sampling algorithms, including the direct Gibbs sampler, the augmented Gibbs sampler, and the interweaving Gibbs sampler. Section~\ref{sec:5:simulation} studies the empirical performance of the proposed methods through simulation experiments and a real financial time-series data analysis. We conclude with a brief discussion.

\section{Preliminaries}\label{sec:2:setup}

{\bf Notation.} We begin by introducing the notations used in the paper. We define $\delta_{c}$ as the degenerate distribution at $c \in \mathbb{R}$. $N(\mu,\sigma^{2};a,b)$ denotes the truncated normal distribution with mean $\mu$ and variance $\sigma^{2}$, restricted to the interval $(a,b)$. $I_A(\cdot)$ denotes the indicator function of a set $A$, which takes the value $1$ on $A$ and $0$ elsewhere. Also, let $\mathcal{S}(p),\, \mathcal{S}^{*}(p)$ and $\mathcal{S}^{**}(p)$ denote the collection of all $p \times p$ symmetric positive definite matrices with non-positive off-diagonal entries, the collection of all $p \times p$ symmetric matrices with positive diagonal and non-positive off-diagonal entries and the collection of all $p \times p$ symmetric strictly diagonally dominant matrices with non-positive off-diagonal entries, respectively. Finally, we write $\|A\|_{F}$ to denote the Frobenius norm of a matrix $A$. We use the standard asymptotic notations $O(\cdot)$ and $o(\cdot)$. Specifically, $f(n)=O(g(n))$ if there exist constants ($C>0$) and $n_0$ such that $|f(n)|\leq C|g(n)|$ for all $n\geq n_0$, whereas $f(n)=o(g(n))$ if $\lim_{n\to\infty}f(n)/g(n)=0$.

\subsection{Generalized likelihood and loss-based Bayesian inference}
\label{sec:2:preliminaries}

In many statistical problems, the full likelihood corresponding to a probabilistic data-generating model may be difficult to evaluate or computationally burdensome to work with, particularly in high-dimensional settings. This has motivated the use of likelihood-type objectives based on loss functions, which serve as tractable surrogates for the full likelihood. The use of such loss-based approaches has a long history in the statistical literature; see, for example, \citep{besag,lindsay,varin2011overview}. These methods are often designed to target the parameter or functional of interest directly, while avoiding the additional complexity associated with specifying a complete probabilistic model for the data.

A key advantage of loss-based formulations is that they can lead to computationally efficient procedures by replacing components of the likelihood that are difficult to evaluate. In the context of Gaussian graphical models, a prominent example is the log-determinant term in the Gaussian likelihood, which incurs $\mathcal{O}(p^3)$ computational cost and requires positive definiteness of the precision matrix at each evaluation. To mitigate this burden, several alternative objectives have been proposed that avoid the log-determinant term while retaining desirable statistical properties. These include methods based on conditional likelihoods and pseudo-likelihoods, such as neighborhood selection \citep{buhlman}, SPACE \citep{Peng}, and SPLICE \citep{rocha2008}, as well as formulations based on surrogate loss functions, including the symmetric lasso \citep{friedman2010applications}, CONCORD \citep{khare2015convex}, quasi-GHS \citep{ZHANG2022154}, CLIME \citep{Cai01062011}, and the D-trace loss \citep{4b5cd03a-e599-3ef7-a1a6-896148b7af15}. These approaches typically reduce computational complexity and can be less sensitive to model misspecification, making them well suited for high-dimensional problems.

In the Bayesian setting, \citet{bissiri2016general} developed a principled framework for incorporating such loss-based objectives. In this approach, the likelihood contribution is replaced by an exponentiated negative loss function, yielding a \emph{generalized posterior} of the form
\[
\pi(\theta \mid \mathbf{Y}) \propto \exp\{-\lambda L(\theta;\mathbf{Y})\} \, \pi(\theta),
\]
where $L(\theta;\mathbf{Y})$ is a chosen loss function, $\pi(\theta)$ is a prior distribution, and $\lambda > 0$ is a scaling parameter controlling the relative influence of the data. This construction provides a coherent updating rule for prior beliefs even when a fully specified likelihood is unavailable. The resulting generalized posterior retains many of the formal properties of Bayesian inference while allowing greater flexibility in modeling and computation.

In this paper, we adopt this perspective and construct a generalized Bayesian model based on the D-trace loss, which yields a computationally tractable posterior formulation while naturally accommodating structural constraints on the precision matrix. The specific form of the generalized likelihood used in our work is described in the sequel. 

\subsection{The D-trace loss} \label{subsec:2.1:dtrace}

We conclude the preliminaries by introducing the D-trace loss and its key properties. The D-trace loss function for estimating $\Omega$ using a $p \times p$ positive definite matrix $\Bar{\Sigma}$ is defined as 
\begin{align} \label{eqn:1:dtrace}
    L_{D}(\Omega,\Bar{\Sigma}) = \frac{1}{2} \, \mathrm{tr}(\Omega^{2} \Bar{\Sigma}) - \mathrm{tr}(\Omega).
\end{align}
The loss function is a difference of two trace terms, which motivates the name \emph{D-trace}. The key properties of $L_{D}$ include the following. First, $L_{D}(\Omega,\cdot)$ is a smooth convex function in $\Omega$, leading to computational advantages. Second, the unique minimizer of $L_{D}(\Omega,\Bar{\Sigma})$ is $(\Bar{\Sigma})^{-1}$, enabling recovery of the precision matrix through minimization of the loss. Motivated by these properties, we employ the generalized likelihood induced by the D-trace loss for estimation of $\Omega$. Finally, the loss is formulated in terms of the covariance matrix (second argument) rather than the precision matrix, allowing direct application in settings where only the covariance matrix is available.

Although the D-trace generalized likelihood shares certain properties with the standard Gaussian likelihood such as convexity and minimization at the inverse of the sample covariance matrix it also offers important advantages. In particular, the $\mathrm{tr}(\Omega)$ term replaces the $\log \det(\Omega)$ term in the Gaussian likelihood, reducing the associated $\mathcal{O}(p^{3})$ computational cost to $\mathcal{O}(p)$. Moreover, evaluation of $\log \det(\Omega)$ requires $\Omega$ to be positive definite, which introduces additional computational overhead in MCMC implementations, as positive definiteness must be enforced at each iteration. This issue is naturally avoided in the D-trace framework. Indeed, in the subsequent section, we relax the positive definiteness constraint on $\Omega$ and allow $\Omega$ to vary in the space of symmetric matrices with positive diagonal entries. Further discussion of this relaxation is provided in Section~\ref{sec:3:model}.

\section{Model Description}\label{sec:3:model}
{\color{black}In this section, we provide a detailed description of the model used in the subsequent sections. Let $\Yi_{1},\ldots,\Yi_{n}$ be a random sample from a $p$-variate mean-zero Gaussian distribution with covariance matrix $\Sigma=\Omega^{-1}$, where $\Omega$ denotes the corresponding precision matrix. We write $\Yb=(\Yi_{1},\Yi_{2},\ldots,\Yi_{n})^{T}$ for the resulting $n\times p$ observation matrix. As described in Section~\ref{sec:1:intro}, our primary objective is to estimate the precision matrix $\Omega=((\omega_{ij}))_{i,j=1}^{p}$ under the totally positive graphical model framework, where the off-diagonal entries of $\Sigma$ are positive. In this setting, the problem reduces to estimating $\Omega$ subject to the $M$-matrix constraint \citep{ying2023adaptiveestimationgraphicalmodels}, namely, $\omega_{ij}\leq 0$ for all $i\neq j$, $i,j\in\{1,\ldots,p\}$. As discussed before, rather than relying on a loss function derived from the strict Gaussian likelihood, we adopt a generalized likelihood based on the D-trace loss introduced in \eqref{eqn:1:dtrace}. }

 In particular, for a data matrix $\Yb=(\Yi_{1},\Yi_{2},\dots,\Yi_{n})^{T}$ of dimension $n \times p$, the generalized likelihood of $\Omega$ based on the D-trace loss can be written as the following.
\begin{align}\label{eqn:2:gen:likelihood}
    L_{D}(\Omega|\Yb)= \exp\left\{n\operatorname{tr}(\Omega)-\frac{n}{2}\operatorname{tr}(\Omega^{2}S)\right\}
\end{align}
where, $S=n^{-1} \Yb^{T}\Yb$ with $(i,j)$-th element $s_{ij}$. Following~\cite{bissiri2016general}, we use the exponentiated negative D-trace loss function as our generalized likelihood. 

Since our goal is to estimate the precision matrix in Bayesian setting, we will now introduce a prior distribution on $\Omega$. We impose independent spike-and-slab priors on the off-diagonal entries of $\Omega$ to induce sparsity, and for the diagonal entries, we introduce iid flat priors. The prior assumptions on the entries of $\Omega$ are as formulated below. For $1 \leq i< j \leq p$, we assume
\begin{equation} \label{eqn:3:prior}
\omega_{ij} \stackrel{iid}{\sim} (1-q) \delta_{0}+q\, \mathcal{N} (0,\tau^{2};-\infty,0), \quad \quad \quad 
\omega_{ii} \propto I_{(0,\infty)}(\omega_{ii})
\end{equation}

where $0<q<1$ and $\tau^{2}>0$ are the hyperparameters denoting the prior probability and the prior variance of the slab respectively.

Note that, the prior is supported on $\mathcal{S}^{*}(p)$, the collection of all $p \times p$ matrices with positive diagonal and non-positive off-diagonal entries, rather than $\mathcal{S}(p)$, the collection of all symmetric positive definite matrices with non-positive off-diagonal entries. This relaxation is feasible due to the structure of the D-trace generalized likelihood, which replaces the log-determinant term with a trace function. It does not create any issues for the task of sparsity selection, but needs to be addressed if positive definite estimates are needed for downstream applications. If a positive definite
estimate of \(\Omega\) is needed for a downstream application, then the
PD relaxation may require an additional post-processing step. In
particular, for a posterior draw or point estimate \(\Omega\) that is not
positive definite, one may project it to the positive definite cone by
shifting its spectrum. Specifically, following the projection idea used in
\citet{jalali2020bconcordscalablebayesian,samanta2022generalized}, one can replace \(\Omega\) by
\[
B(\Omega)
=
\begin{cases}
\Omega, & \text{if } \Omega \text{ is positive definite},\\
\Omega-\lambda_{\min}(\Omega)I_p+\epsilon I_p,
& \text{if } \lambda_{\min}(\Omega)\le 0,
\end{cases}
\]
where $\lambda_{\min}(\Omega)$ is the minimum eigenvalue of $\Omega$ and \(\epsilon>0\) is a small user-specified constant. This
transformation leaves positive definite matrices unchanged and shifts any
non-positive-definite matrix into the positive definite cone by increasing
its diagonal entries. However, our empirical results under various
settings show that a substantial proportion of posterior draws of
\(\Omega\) from our proposed samplers in Section \ref{sec:4:methods} are already positive definite (see Supplemental Section~\ref{sec:sup:PD}). 
If we further restrict $\mathcal{S}^{*}(p)$ to another class, say 
$\mathcal{S}^{**}(p)$, consisting of all $p\times p$ symmetric matrices 
with positive diagonal entries, non-positive off-diagonal entries, and strict 
diagonal dominance, that is,
\[
\mathcal{S}^{**}(p)
=
\left\{
\Omega\in\mathcal{S}^{*}(p):
\omega_{ii}>\sum_{j\neq i}|\omega_{ij}|,\quad i=1,\ldots,p
\right\},
\]
then, every matrix 
in $\mathcal{S}^{**}(p)$ is positive definite. Therefore, restricting the parameter space from $\mathcal{S}^{*}(p)$ to 
$\mathcal{S}^{**}(p)$ provides an alternative way to obtain positive definite 
posterior draws without requiring any additional post-processing 
or projection onto the positive definite cone. Such formulations are common in 
the literature; see, for example, 
\cite{ying2023adaptiveestimationgraphicalmodels,article}. Due to space 
constraints, a detailed analysis of this restricted formulation, including the 
corresponding sampling algorithm and additional simulation results, is provided 
in the Supplementary Material.


\subsection{The (Generalized) Posterior Distribution} \label{subsec:3.1:fullpost}

With the generalized likelihood and the specified prior in place, we now derive the joint generalized posterior distribution of $\Omega$. A direct calculation yields the following lemma.

\begin{lemma}[Joint posterior kernel]\label{lemma:1:post:jt}
Let $S=n^{-1}\Yb^{T}\Yb$. Under the D-trace generalized likelihood in~\eqref{eqn:2:gen:likelihood} and the spike-and-slab prior in~\eqref{eqn:3:prior}, the joint generalized posterior density of $\Omega$, given the data matrix $\Yb$, is determined, up to a normalizing constant, by the kernel
\begin{equation}\label{eqn:4:post:exact}
\fontsize{11}{12}\selectfont
\pi_D(\Omega\mid \Yb)
\propto
\exp\left\{
n\sum_{i=1}^{p}\omega_{ii}
-\frac{n}{2}\operatorname{tr}(\Omega^2 S)
\right\}
\prod_{\substack{i,j=1\\ i<j}}^{p}
\left[
(1-q)I_{\{0\}}(\omega_{ij})
+
q\frac{2}{\tau\sqrt{2\pi}}
\exp\left\{-\frac{\omega_{ij}^{2}}{2\tau^{2}}\right\}
\right]I_{ \mathcal{S}^{*}(p)}(\Omega). 
\end{equation}
\end{lemma}
It is important to note that the generalized posterior density in \eqref{eqn:4:post:exact} is not induced by an ordinary likelihood arising from a fully specified probability model. Rather, the generalized likelihood is obtained by exponentiating a loss function. In addition, the priors imposed on the model parameters are not all proper. Consequently, the propriety of the resulting generalized joint posterior  is not automatic. Equivalently, the finiteness of the normalizing constant associated with the kernel in Lemma~\ref{lemma:1:post:jt} must be verified.
Examples of improper priors leading to improper posteriors are discussed in \cite{dixit2021posteriorimproprietysparsebayesian}. The implications of posterior impropriety for Gibbs samplers have been studied in \cite{6621c4dcb35a461e933fb952ac6dbd72}, \cite{1e5ec504-73f2-32f2-91a6-4be5574211d8}, and \cite{25003dfe-dd6d-37e1-a5d7-dd568a8c251c}. In particular, it has been shown that an MCMC output may fail to exhibit any obvious numerical instability or unreasonable behavior even when the corresponding posterior is improper~\cite{1e5ec504-73f2-32f2-91a6-4be5574211d8}; moreover, standard MCMC-based estimates may converge to zero with probability one under posterior impropriety. These considerations make it essential to rigorously establish the propriety of the generalized posterior density $\pi_D(\cdot \mid {\bf Y})$. 

The following lemma addresses this issue. It shows that, although the construction does not start from an exact likelihood and the prior specification includes improper components, the posterior distribution stated in Lemma~\ref{lemma:1:post:jt} is a proper probability distribution. This result provides the foundation for the coherence of the subsequent posterior analysis.

\begin{lemma}[Posterior propriety]\label{lemma:2:post:proper}
Suppose that $s_{ii}>0,\ i=1,\ldots,p$. Then, for any sample size $n$ and ambient dimension $p$, the joint generalized posterior density of $\Omega$, given $\Yb$, is proper. Equivalently, the normalizing constant associated with the posterior kernel in Lemma~\ref{lemma:1:post:jt} is finite. 
\end{lemma}

The condition \(s_{ii}>0\), \(i=1,\ldots,p\), in Lemma~\ref{lemma:2:post:proper} is mild. Since \(S=n^{-1}\Yb^{T}\Yb\), \(s_{ii}=0\) if and only if the \(i\)th coordinate is zero for all observations. Thus, it only excludes coordinates with zero empirical second moment, an event of probability zero for continuous variables with no atom at zero. The proof is given in Supplementary Section~\ref{app:A1:pf:l2}. Consequently, the posterior is proper and posterior inference is well-defined in the usual probabilistic sense.

\section{Posterior Computation and Sampling} \label{sec:4:methods}
Note that the form of the joint (generalized) posterior density $\pi_{D}(\Omega|\Yb)$ in~\eqref{eqn:4:post:exact} is not tractable for closed-form analytical evaluation or direct sampling. Therefore, in this section, we introduce three MCMC methods to sample from the posterior. 

\subsection{Componentwise (Direct) Gibbs Sampler}\label{subsec:4.1:directgibbs}
We consider a componentwise Gibbs sampler to sample from $\pi_{D}(\Omega|\Yb)$. To this end, we compute the conditional posterior densities of the entries of $\Omega$, given the remaining entries. Straightforward calculations yield the following conditional posteriors. 
\begin{lemma} \label{lemma:3:post:comp}
The conditional posterior distributions of the off-diagonal entries of $\Omega$, given the remaining entries, are obtained as follows. For $1 \leq i < j \leq p$,
\begin{align}\label{eqn:5:fullcond:offdiag}
    \omega_{ij}\mid \Yb, \Omega_{-(ij)} 
    \sim (1 - q_{ij}) \delta_0 
    + q_{ij}\, \mathcal{N}\!\left(-\frac{b_{ij}}{a_{ij}},\, \frac{1}{n a_{ij}};\, -\infty, 0 \right),
\end{align}
where
\[
q_{ij} = \frac{c_{ij}}{1 + c_{ij}}, \quad 
c_{ij} = \frac{2q}{(1 - q)\sqrt{n a_{ij} \tau^2}} \exp\!\left(\frac{n b_{ij}^2}{2 a_{ij}}\right)\Phi\left(
b_{ij}\sqrt{\frac{n}{a_{ij}}}
\right),
\]
\[
a_{ij} = s_{ii} + s_{jj} + \frac{1}{n \tau^2}, \quad 
b_{ij} = \sum_{j' \neq j} \omega_{i j'} s_{j j'} + \sum_{i' \neq i} \omega_{i' j} s_{i i'},
\]
and $s_{ij}$ denotes the $(i,j)$-th entry of $S = \frac{1}{n}\Yb^\top \Yb$. For any matrix $A$, $A_{-(ij)}$ denotes the vector of all the elements in the upper triangle except the $(i,j)$-th entry.

The conditional posterior distributions of the diagonal entries of $\Omega$, given the remaining entries, are given by
\begin{align}\label{eqn:6:fullcond:diag}
    \omega_{ii}\mid \Yb, \Omega_{-(ii)} 
    \sim \mathcal{N}\!\left(\frac{1 - d_i}{s_{ii}},\, \frac{1}{n s_{ii}};\, 0, \infty \right),
\end{align}
where $d_i = \sum_{i' \neq i} \omega_{i i'} s_{i i'}, 
\quad \text{for } i = 1, 2, \dots, p.$
    
\end{lemma}

{\color{black}
\begin{algorithm}[!h]
	\caption{Direct Gibbs Sampler} \label{algo:1:directgibbs} 
    {\setstretch{1.22}
	\begin{algorithmic}[1]
	\Procedure{D-GIBBS($\Omega^{init},\Yb, T, q, \tau^2$)}{}
        \State $\Omega^{(0)}=\Omega^{init}$
		\State $S = n^{-1}\Yb^{T} \Yb$
        \For{$iter=1,2,\dots,T$}
          \State $\Omega^{new} = \Omega^{(iter-1)}$
		\For {$i=1,2,\ldots,p-1$} \Comment{Updating off-diagonals of $\Omega$}
		   \For {$j=i+1,\ldots,p$}

	        	\State $a \gets s_{ii} + s_{jj} + 1/n\tau^{2},\,b \gets (\Omega^{new}_{.i})^TS_{.j}+(\Omega_{.j}^{new})^TS_{.i} - (s_{ii}+s_{jj}) \omega_{ij}^{new}$
		        \State $P(0) \gets 1, \, P(1) \gets \sqrt{\frac{1}{na\tau^2}}\frac{2q}{1-q}\exp{\left[\frac{nb^2}{2a}\right]} \Phi\left(
b\sqrt{\frac{n}{a}}\right)$
				\If {$P(1) = \infty$}
				\State $\omega^{new}_{ij} \gets \mathcal{N}\left(-\frac{b}{a},\frac{1}{na};-\infty,0\right),\,\omega_{ji}^{new} \gets \omega_{ij}^{new}$
				\Else 
				\State $P \gets {P/\text{sum}(P)}$
				  \State $\omega_{ij}^{new} \sim P(0) \delta_0 + P(1) \mathcal{N}\left(-\frac{b}{a},\frac{1}{na};-\infty,0\right),\, \omega_{ji}^{new} \gets \omega_{ij}^{new}$ 
                \EndIf
		    \EndFor 
		\State $d \gets 1-(\Omega_{.i}^{new})^{T} S_{.i}+\omega_{ii}^{new}s_{ii} $ \Comment{Updating diagonals of $\Omega$}
		\State $\omega_{ii}^{new} \gets \mathcal{N}(\frac{d}{s_{ii}},\frac{1}{ns_{ii}};0,\infty)$ 
		\EndFor
		\State Repeat Steps 18 - 19 for $i = p$
        \State $\Omega^{(iter)} \gets \Omega^{new}$
         \EndFor
		\State \textbf{return} $\Omega^{(1)},\Omega^{(2)}, \dots,\Omega^{(T)}$
	\EndProcedure
	\end{algorithmic}
    }
\end{algorithm}
}

The proof of Lemma~\ref{lemma:3:post:comp} is provided in Supplementary Section~\ref{app:A2:pf:l3}. Posterior samples can be generated using a Gibbs sampler based on these conditional distributions. A schematic of this sampler, which we refer to as the {\em Direct Gibbs Sampler}, or in short, {\em D-Gibbs} is outlined in Algorithm~\ref{algo:1:directgibbs} below. Our next lemma focuses on the computational complexity of this 
Direct Gibbs Sampler.

\begin{lemma}\label{lemma:4:compDirect}
    The computational complexity of each iteration of the direct Gibbs sampler in Algorithm~\ref{algo:1:directgibbs} is $\mathcal{O}(p^{3})$.
\end{lemma}

The proof of Lemma~\ref{lemma:4:compDirect} is provided in Supplementary Section~\ref{app:A3:pf:l4}. Note that the $\mathcal{O}(p^{3})$ computational complexity in Lemma~\ref{lemma:4:compDirect} is comparable to that of inverting a $p \times p$ matrix. This results in substantial computational overhead, posing a significant limitation for implementing this sampler in high dimensions, particularly when $p \gg n$. Moreover, the dependence induced by the direct Gibbs updates may lead to inefficient MCMC estimates. In the following section, we address these issues—especially the computational bottleneck—by developing an alternative Gibbs sampler based on an augmented framework. 

\subsection{Augmented Gibbs Sampler} \label{subsec:4.2:aug:gibbs}
Note that the non-zero correlation between the entries of $\Omega$ restricts joint sampling of all elements, thus enforcing sequential one-by-one sampling, which contributes to the computational overhead. To address this issue, we develop an Augmented Gibbs Sampler framework by generalizing the {\em Anti-correlation Gaussian} approach proposed in~\cite{zheng2023gibbs} to matrix-variate settings. The goal is to introduce latent variables such that the posterior distributions of the entries of $\Omega$ are conditionally independent given these variables, thereby enabling efficient joint sampling. The resulting augmented posterior must, by construction, remain consistent with the original posterior distribution. A computationally efficient generalization of the vector-based framework in \cite{zheng2023gibbs} to the more involved matrix-variate setting necessitates several nontrivial and conceptually novel ideas, which we present in detail below. 

To motivate the augmented formulation, we introduce a matrix-variate latent variable $R\in\mathbb{R}^{p\times p}$ such that
\begin{align}\label{eqn:7:distR}
    R\mid \Yb,\Omega \sim \mathcal{MN}_{p\times p}\bigl (\Omega(kI_p-nS),\,I_p,\,kI_p-nS\bigr),
\end{align}
where $k$ is chosen sufficiently large so that $kI_p-nS$ is positive definite. Equivalently, conditional on $\Yb$ and $\Omega$, the rows of $R$ are independent and
\[
    R_{i\cdot}\mid \Yb,\Omega
    \sim \mathcal{N}_p\bigl(\{\Omega(kI_p-nS)\}_{i\cdot},\,kI_p-nS\bigr),
    \qquad i=1,\ldots,p.
\]
The proposed augmentation in \eqref{eqn:7:distR} is valid since \(R\) is
introduced through a \textit{proper} matrix-variate normal distribution
given \(\Omega\) and \(\Yb\). Indeed, by Lemma~\ref{lemma:2:post:proper},
the original generalized posterior for \(\Omega\) is \textit{proper}.
Moreover, for each fixed \(\Omega\) and \(\Yb\), the conditional
distribution of \(R\) has total mass one. Consequently, the augmented
joint posterior of \((R,\Omega)\) given \(\Yb\) is also proper. In
particular, integrating the augmented posterior with respect to \(R\) and
then \(\Omega\) yields one, while integrating with respect to \(R\) alone
recovers the original generalized posterior for \(\Omega\). Next we'll show this augmentation yields a conditionally independent posterior structure for the elements of $\Omega$, stated in the following lemma.

\begin{lemma}[Augmented conditional posterior distribution of $\Omega$]\label{lemma:5:augcond}
Under the augmented formulation in \eqref{eqn:7:distR}, the conditional posterior densities of the entries of $\Omega$ given $R$ and $\Yb$ are as follows. For $1\le i<j\le p$,
\begin{align}\label{eqn:8:augcond}
    \omega_{ij}\mid R,\Yb
    &\sim (1-q_{ij}^{*})\delta_{0}
    +q_{ij}^{*}\mathcal{N}\left(\frac{R_{ij}+R_{ji}}{k^{*}},\frac{1}{k^{*}};-\infty,0\right), \nonumber\\
    \omega_{ii}\mid R,\Yb
    &\sim \mathcal{N}\left(\frac{R_{ii}+n}{k},\frac{1}{k};0,\infty\right),
\end{align}
where
\[
q_{ij}^{*}=\frac{c_{ij}}{1+c_{ij}},\qquad
c_{ij}=\frac{2q}{1-q}\frac{1}{\tau\sqrt{k^{*}}}
\exp\left\{\frac{(R_{ij}+R_{ji})^{2}}{2k^{*}}\right\}\Phi\left(
-\frac{R_{ij}+R_{ji}}{\sqrt{k^*}}
\right),
\qquad
k^{*}=2k+\frac{1}{\tau^{2}}.
\]
Moreover, conditional on $R$ and $\Yb$, all entries in the upper triangle of $\Omega$ are mutually independent under the augmented posterior.
\end{lemma}

The proof of this Lemma is provided in Supplementary Section~\ref{app:A4:pf:l5}. It follows immediately from Lemma~\ref{lemma:5:augcond} that, conditional on the augmented variable $R$, sampling from the posterior distribution of $\Omega$ can be carried out in $\mathcal{O}(p^2)$ operations. However, at each iteration of the Gibbs sampler, one must also sample $R$ conditional on the current value of $\Omega$ from the matrix-normal distribution in \eqref{eqn:7:distR}. Direct sampling from this distribution requires computing a matrix square root, or equivalently a Cholesky factor, of the $p\times p$ matrix $kI_p-nS$, which entails $\mathcal{O}(p^3)$ computational cost. Thus, despite the conditional independence in the augmented posterior, the direct implementation has the same cubic-order bottleneck as the naive elementwise Gibbs sampler described in Section~\ref{subsec:4.1:directgibbs}. As mentioned before this becomes particularly restrictive in high-dimensional settings where $n\ll p$.

However, we leverage specific structure of the matrix \(kI_p-nS\) to develop a significantly more efficient alternative. In particular, if \(\Yb\) denotes the \(n\times p\) data matrix, then \(S=n^{-1}\Yb^{T}\Yb\), and hence \(kI_p-nS=kI_p-\Yb^{T}\Yb\). Exploiting a low-rank structure to reduce the computational burden is common in the literature, especially in high-dimensional settings where \(p \gg n\). For example, in the regression context, \citet{Bhattacharya} consider sampling from a Gaussian distribution with covariance matrix \(\left(X^{T}\Omega X+D\right)^{-1}\), where \(D\) is a \(p\times p\) positive definite matrix and \(X\) is an \(n\times p\) design matrix. They exploit the Woodbury matrix identity, which reduces the computation to the inversion of an \(n\times n\) matrix instead of a \(p\times p\) matrix when \(p \gg n\). In a related setting, \citet{Nishimura02102023} use a preconditioned conjugate gradient method to iteratively solve the corresponding linear system.

\noindent Our present setting differs from the aforementioned regression examples in two important respects. First, the covariance structure involves the matrix difference \(kI_p-nS\), rather than a sum of a positive definite matrix and a low-rank term. Second, the augmented object to be sampled is matrix-valued rather than vector-valued. A direct, but computationally inefficient, approach would be to apply the sampling scheme of~\citet[Theorem~2.1]{zheng2023gibbs} separately to each row of \(R\), thereby repeating the same procedure \(p\) times. As discussed in \cite{zheng2023gibbs}, generating a sample for each row of \(R\) has computational complexity \(\mathcal{O}(p^2+n^2)\). Therefore, repeating this procedure \(p\) times leads to an overall complexity of \(\mathcal{O}(p^3+n^2p)\), which remains of order \(\mathcal{O}(p^3)\) when \(n\ll p\). This inefficiency arises because the procedure treats the rows separately and does not exploit the matrix-normal structure of \(R\). To overcome this obstacle, we propose a novel and efficient sampling scheme for matrix-normal distributions of the form \eqref{eqn:7:distR}. The following lemma formalizes this scheme.
\begin{lemma}[Fast matrix-normal sampling under Gram structure]\label{lemma:6:struct:gauss}
Let $U\in\mathbb{R}^{p\times n}$ satisfy $nS=kUU^{T}$, where $k$ is chosen such that $I_{n}-U^{T}U$, equivalently $kI_p-nS$, is positive definite. Let $A\in\mathbb{R}^{n\times n}$ denote the Cholesky factor of $I_{n}-U^{T}U$, so that $AA^{T}=I_{n}-U^{T}U$. Let $V_{1}\in\mathbb{R}^{p\times p}$ and $V_{2}\in\mathbb{R}^{p\times n}$ be independent random matrices with i.i.d.\ standard normal entries. Define
\[
X=\sqrt{k}\left\{V_{1}(I_{p}-UU^{T})+V_{2}A^{T}U^{T}\right\}.
\]
Then $X\sim \mathcal{MN}_{p\times p}\bigl(0,I_{p},kI_{p}-nS\bigr)$.
Moreover, after precomputing $A$, one draw of $X$ can be generated in $\mathcal{O}(p^{2}n+pn^{2})$ operations, which reduces to $\mathcal{O}(p^{2}n)$ when $n\ll p$.
\end{lemma}
The proof of Lemma~\ref{lemma:6:struct:gauss} is provided in Supplementary Section~\ref{app:A5:pf:l6}. In particular, Lemma~\ref{lemma:6:struct:gauss} can be used to sample from \eqref{eqn:7:distR} by taking $U=k^{-1/2}\Yb^{T}$, so that $kUU^{T}=\Yb^{T}\Yb=nS$. The key advantage of Lemma~\ref{lemma:6:struct:gauss} is that the Cholesky factor $A$ can be computed once before running the MCMC chain, since $S$ is fixed after observing the data. This preprocessing step requires $\mathcal{O}(pn^{2}+n^{3})$ operations. Conditional on this preprocessing, one draw from the distribution can be generated at each MCMC iteration in $\mathcal{O}(p^{2}n+pn^{2})$ operations using only matrix multiplications. In particular, when $n \ll p$, the per-draw complexity reduces to $\mathcal{O}(p^{2}n)$. {\color{black} For $n>p$, we can sample from the matrix normal directly with $O(p^3)$ complexity.}

In the general Bayesian covariance matrix estimation literature using shrinkage priors (even without sign constraints), under the exact Gaussian likelihood, the time complexity of the sampling algorithms is typically $\mathcal{O}(p^{4})$ when $n>p$. Recently, \cite{gao2026ordermagnitudetimecomplexity} reduce this cost to $\mathcal{O}(p^{3})$ in the high-dimensional regime $n=o(\sqrt p)$. In contrast, our per-draw complexity is $\mathcal{O}(p^{2}n)$, which becomes $\mathcal{O}(p^{2.5})$ when $n=o(\sqrt{p})$, and is therefore lower in this regime. Thus, even without sign constraints, our approach of employing the D-Trace generalized likelihood and relaxing the positive-definiteness constraint can substantially improve computational scalability compared to existing methods. The quadratic nature of the D-trace loss, relaxation of PD constraint, and the conditional independence entries of $\Omega$ given $R$ also enable seamless integration of any sign constraint on $\Omega$ in our framework, which is not the case with the Gaussian likelihood-based method. Hence, the computation scalability issue for this method becomes even more acute under the sign constraints. Moreover, the empirical results in Section~\ref{sec:5:simulation}
demonstrate that this computational advantage is not obtained at the cost
of any systematic loss of estimation accuracy.

A remaining practical issue is the choice of $k$ in the proposed algorithm. Since the augmentation requires $kI_p-nS$ to be positive definite, it is enough to choose $k>n\lambda_{\max}(S)$. We set $k=\{n\epsilon \lambda_{\max}(S)\}$, where $\epsilon=1.001$. Since $S$ is fixed after observing the data, this quantity can be computed once before starting the MCMC chain. A direct computation requires $\mathcal{O}(p^3)$ operations, which is a one-time preprocessing cost and does not affect the per-iteration complexity of the sampler.
Combining Lemmas~\ref{lemma:5:augcond} and~\ref{lemma:6:struct:gauss}, we now present the entrywise Gibbs sampler in Algorithm~\ref{algo:2:auggibbs}.

\begin{algorithm}[!ht]
	\caption{Augmented Gibbs Sampler} \label{algo:2:auggibbs} 
    {\setstretch{1.22}
	\begin{algorithmic}[1]
	\Procedure{A-GIBBS($\Omega^{init},\Yb,T,\varepsilon, q, \tau^2$)}{}
        \State $\Omega^{(0)} = \Omega^{init}$
		\State $S = n^{-1}\Yb^{T} \Yb$, $\lambda=$ Maximum eigenvalue of $nS$ 
        \State $k \gets \varepsilon \lambda,\,k^{*} \gets 2k+\tau^{-2}$
        \For{iter $=1,2,\dots,T$}
        \If {$p> n$}  \Comment{Sampling Augmented variable $R$}
        \State $R \gets \mathcal{MN}_{p \times p}\bigl(\Omega^{(iter-1)}(kI_p-nS),\,I_p,\,kI_p-nS\bigr)$ using Lemma~\ref{lemma:6:struct:gauss}
        \Else 
        \State $R \gets \mathcal{MN}_{p \times p}\bigl(\Omega^{(iter-1)}(kI_p-nS),\,I_p,\,kI_p-nS\bigr)$ directly
        \EndIf
        \State $\Omega^{new} = \Omega^{(iter-1)}$
		\For {$i=1,2,\ldots,p-1$} \Comment{Updating off-diagonals of $\Omega$}
		   \For {$j=i+1,\ldots,p$}
		        \State $P(0) \gets 1, P(1) \gets \frac{2q}{1-q}\frac{1}{\tau\sqrt{k^{*}}}
\exp\left\{\frac{(R_{ij}+R_{ji})^{2}}{2k^{*}}\right\}\Phi\left(
-\frac{R_{ij}+R_{ji}}{\sqrt{k^*}}\right)$
				\If {$P(1) = \infty$}
				\State $\omega_{ij}^{new} \gets \mathcal{N}\left(\frac{R_{ij}+R_{ji}}{k^{*}},\frac{1}{k^{*}};-\infty,0\right),\, \, \omega_{ji}^{new} \gets \omega_{ij}^{new}$
				\Else 
				\State $P \gets {P/\text{sum}(P)}$ 
                \State $\omega_{ij}^{new} \sim P(0) \delta_0 + P(1) \mathcal{N}\left(\frac{R_{ij}+R_{ji}}{k^{*}},\frac{1}{k^{*}};-\infty,0\right),\, \omega_{ji}^{new} \gets \omega_{ij}^{new}$
				\EndIf
		    \EndFor 
		\State $\omega_{ii}^{new} \gets \mathcal{N}(\frac{R_{ii}+n}{k},\frac{1}{k};0,\infty)$  \Comment{Updating diagonals of $\Omega$}
		\EndFor
		\State Repeat Step 21 for $i = p$
		\State $\Omega^{(iter)} \gets \Omega^{new}$
        \EndFor
        \State \textbf{return} $\Omega^{(1)},\Omega^{(2)},\dots,\Omega^{(T)}$ 
	\EndProcedure	
	\end{algorithmic}
    }
\end{algorithm}

\noindent Note that, in Algorithm~\ref{algo:2:auggibbs}, the entries of \(\Omega\) are conditionally independent given the augmented variable \(R\). Hence each entry of $\Omega$ (given $R$) can now be independently sampled with $O(1)$ computations requiring a total of $O(p^2)$ computations for sampling \(\Omega\) (this can be further improved through parallel sampling, if resources are available). The discussion on the computational complexity of the augmented A-Gibbs sampler in the high-dimensional regime is formalized in the following lemma.

\begin{lemma}\label{lemma:7:compaug}
The computational complexity of a single iteration of the
proposed augmented Gibbs sampler in Algorithm~\ref{algo:2:auggibbs} is
\(\mathcal{O}(\min\{p^2n,p^3\})\). When $p \gg n$, this complexity reduces to $O(np^2)$, which is an order of magnitude smaller than the complexity of the direct Gibbs sampler.
\end{lemma}


\subsection{The Interweaving Algorithm} \label{subsec:4.3:interweaving}

\noindent
As described in Section~\ref{subsec:4.1:directgibbs}, the componentwise direct Gibbs sampler suffers from substantial computational overhead, with per-iteration complexity of $\mathcal{O}(p^3)$, and induces strong dependence among updates. Together, these factors can lead to slow mixing and inefficient MCMC estimation, particularly in high-dimensional settings. To address these issues, we introduced a data augmentation framework in Section~\ref{subsec:4.2:aug:gibbs} and proposed an augmented Gibbs sampler.

Although the augmented sampler achieves significant computational gains, exhibits substantially improved mixing, and consistently delivers superior sparsity recovery in our empirical studies (see Sections~\ref{sec:5:simulation} and \ref{subsec:5.2:DataAnalysis}), it operates in an expanded parameter space through the introduction of latent variables. In contrast, the direct Gibbs sampler updates the entries of $\Omega$ under the original posterior parameterization, thereby preserving the intrinsic dependence structure of the model.

These two samplers therefore correspond to different representations of the same posterior distribution, each inducing a distinct Markov chain geometry. The augmented sampler leverages conditional independence to enable efficient global exploration, which translates into improved sparsity selection, as observed in both our simulation studies and real data analysis. On the other hand, the direct sampler performs local updates in the original parameter space and can provide complementary refinements of the chain, although by itself it exhibits slower mixing and comparatively poorer sparsity recovery in practice.

Motivated by these observations, we propose an interweaving algorithm that combines the strengths of both sampling schemes. The proposed sampler is designed to improve exploration of the parameter space both across different sparsity patterns and within a fixed sparsity pattern. The high-level idea behind this interweaving strategy is inspired by the ancillary-sufficiency interweaving strategy of~\cite{Yu01012011}. In particular, we alternate between the augmented and direct representations of the posterior, thereby combining the rapid global exploration afforded by the augmented sampler with the corrective local updates of the direct sampler. The algorithm proceeds by iteratively using the augmented sampler and invoking the direct sampler at every \(N\)-th iteration, for some pre-specified positive integer \(N\), to encourage exploration across distinct sparsity patterns. Our empirical choice of \(N\) is specified in Section~\ref{sec:5:simulation}. The resulting interweaving strategy is described in Algorithm~\ref{algo:3:interweaving}.
\begin{algorithm}[!ht]
{\setstretch{1.22}
	\caption{Interweaving Gibbs Sampler} \label{algo:3:interweaving} 
	\begin{algorithmic}[1]
	\Procedure{I-GIBBS($\Omega^{init},\Yb,T,\varepsilon,q,\tau^2,N$)}{}
        \State $\Omega^{(0)}=\Omega^{init}$
		\State $S = n^{-1}\Yb^{T} \Yb$, $\lambda=$ Maximum eigenvalue of $nS$ 
        \For{$iter =1,2,\dots,T$}
          \If{\(\mathrm{iter}\equiv 0 \pmod N\)}
               \State Perform Steps 5-21 of \textsc{D-GIBBS} in Algorithm~\ref{algo:1:directgibbs} with $\Omega^{init} = \Omega^{(iter-1)}$
            \Else
               \State Perform Steps 6-25 of \textsc{A-GIBBS} in Algorithm~\ref{algo:2:auggibbs} with $\Omega^{init} = \Omega^{(iter-1)}$
            \EndIf
            \State $\Omega^{(iter)} \gets$ sampled matrix in Steps 5-9
        \EndFor
        \State \textbf{return} $\Omega^{(1)},\Omega^{(2)},\dots,\Omega^{(T)}$ 
	\EndProcedure	
	\end{algorithmic}
    }
\end{algorithm}
Empirically, we find that while the augmented sampler already performs strongly, incorporating occasional direct updates further enhances stability and overall mixing without compromising computational efficiency. 

\subsection{Choice of Hyperparameters} \label{subsec:4.4:hyper}
\textcolor{black}{The selection of hyperparameters is an important aspect of Bayesian modeling. Although cross-validation is widely used for hyperparameter tuning, it can be computationally prohibitive in high-dimensional settings with large \(p\). We therefore adopt standard choices from the literature for \(q\) and \(\tau^{2}\). Specifically, motivated by \cite{Narisetty_he,samanta2022generalized}, \(q\) is selected through a grid search over \(\{1/p,1/p^{1.5},1/p^2\}\)}.

For the hyperparameter \(\tau^{2}\), one may either fix it at a value close to \(1\) or adopt a fully Bayesian specification by assigning an objective Inverse-Gamma prior with shape parameter \(10^{-4}\) and rate parameter \(10^{-8}\); see \cite{blasso,samanta2022generalized}.
 If a fully Bayesian approach is employed, some modifications are necessary to Algorithms~\ref{algo:1:directgibbs},~\ref{algo:2:auggibbs} and~\ref{algo:3:interweaving}. First, posterior propriety is not guaranteed in this setting if $n < p$ and improper priors are used for the diagonal entries of $\Omega$. Hence, when $n < p$, we employ independent diffuse Gaussian priors with mean $0$ and precision parameter $0.01$ for each diagonal entry of $\Omega$. As a result, Step 18 of Algorithm \ref{algo:1:directgibbs} will change to 
$$
\omega_{ii} \gets \mathcal{N} \left( \frac{nd}{ns_{ii}+0.01},\frac{1}{ns_{ii}+0.01};0,\infty \right), 
$$

\noindent
and Step 22 of Algorithm \ref{algo:2:auggibbs} will change to 
$$
\omega_{ii} \gets \mathcal{N} \left( \frac{R_{ii}+n}{k+0.01},\frac{1}{k+0.01};0,\infty \right). 
$$

\noindent
Finally, the following Inverse-Gamma update for $\tau^2$ is needed in each iteration of the various Gibbs samplers. 
$$
\tau^2 \sim \mbox{Inverse-Gamma}(10^{-4}+0.5 \displaystyle \sum_{i<j} I_{\mathbb{R} \setminus \{0\}} (\omega_{ij}),10^{-8}+0.5 \sum_{i<j}\omega_{ij}^2)
$$

\subsection{Sparsity Selection and Estimation based on Gibbs Output}\label{subsec:4.5:final_estimation}
Suppose that we have the output sequence from implementing one of the three algorithms proposed above. After discarding the first $B$ samples as burn-in, where $B<T$ is a suitable pre-defined number, we can use the remaining $T^{*} = T-B$ samples to estimate the graph structure and the corresponding precision matrix. Let us write the post burn-in samples as $\{\bar{\Omega}^{(t)}\}_{t=1}^{T^{*}}$. To estimate the sparsity pattern of the precision matrix, we adapt the majority voting approach~\citep{Barbieri_2004}. For this, we compute the proportion of non-zero outputs out of the $T^{*}$ final samples for each of the off-diagonal entries as given below,
$$\hat{q}_{ij} = \frac{1}{T^{*}} \sum_{t=1}^{T^{*}} I_{\mathbb{R} \backslash \{0\}}(\bar{\omega}_{ij}^{(t)})$$
where $\bar{\omega}_{ij}^{(t)} $ is the $(i,j)$-th entry of $\bar{\Omega}^{(t)}$. According to the majority voting approach, we include an edge in the graph structure if the corresponding $\hat{q}_{ij}$ is more than $1/2$. Alternatively, we estimate the sparsity pattern $\eta_{ij} = I(\omega_{ij} \neq 0)$ of the precision matrix $\Omega$ as follows.
\begin{align*}
    \hat{\eta}_{ij} =  
    \begin{cases}
    1, & \text{if } \hat{q}_{ij} \geq \frac{1}{2} \\ 
    0, & \text{otherwise.} 
    \end{cases}
\end{align*}
Once the sparsity pattern is determined, the non-zero entries of $\Omega$ are estimated by simply taking the average of the corresponding entry over the samples where that particular entry is non-zero. More specifically, we compute the entries of the final estimate $\widehat{\Omega}$
\begin{align*}
\widehat{\omega}_{ij} = \begin{cases} \dfrac{\sum_{t=1}^{T^*} \bar{\omega}_{ij}^{(t)} I(\bar{\omega}_{ij}^{(t)} \neq 0)}{\sum_{t=1}^{T^*} I(\bar{\omega}_{ij}^{(t)} \neq 0)}, & \text{if } \hat{\eta}_{ij} = 1 \\[10pt] 
\phantom{\dfrac{\sum_{t=1}^{T^*}}{\sum_{t=1}^{T^*}}} 0, & \text{otherwise.} 
\end{cases}    
\end{align*}

Alternatively, one can simply consider the average of all the $T^{*}$ outputs as an estimate of the non-zero entries. Entrywise credible intervals for the non-zero entries of $\Omega$ can also be constructed using the corresponding Gibbs outputs as an observed sample.


\section{Numerical Experiments}\label{sec:5:simulation}
We implement the proposed algorithms in various settings to analyze both synthetic and real-world data and compare the sparsity selection and estimation performances of the proposed methods with the existing benchmark methods. Relevant approaches include GLASSO~\citep{glasso2009}, CLIME~\citep{Cai01062011}, GGL~\citep{article} and SLTP~\citep{wang2020learninghighdimensionalgaussiangraphical}. \textcolor{black}{Among these, the first two methods estimate graphs in general settings, whereas GGL and SLTP are designed specifically for the \(M\)-matrix constraint. GGL estimates a sparse precision matrix, but SLTP targets only the graph structure. We omit SLTP from the reported tables because it does not provide a well-defined final precision matrix estimate and, more critically, scales poorly with sample size and dimension, with some larger runs failing to terminate within a reasonable time.} As specified in Algorithms~\ref{algo:1:directgibbs},~\ref{algo:2:auggibbs} and~\ref{algo:3:interweaving}, we will refer to the three proposed samplers as D-Gibbs, A-Gibbs, and I-Gibbs, respectively, where D-Gibbs denotes the Componentwise Direct Gibbs Sampler, A-Gibbs denotes the Augmented Gibbs Sampler, and I-Gibbs denotes the Interweaving Direct-Augmented Gibbs Sampler. For the proposed methods, hyperparameters are chosen based on recommendations in Section \ref{subsec:4.4:hyper}. For GLASSO, GGL and CLIME, the respective hyperparameters are chosen based on cross-validation over author-recommended ranges. For I-Gibbs, we take $N=100$, i.e., at every $100$-th iteration, we invoke one iteration of the D-Gibbs algorithm.  

\subsection{Synthetic Data Analysis} \label{subsec:5.1:simulation}

\begin{figure}[h]
    \centering
    \includegraphics[width=\textwidth, height = 0.18\textheight]{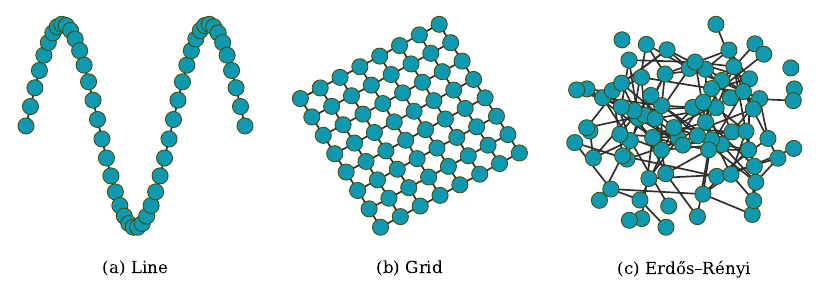}
    \caption{Illustrations of graph structures used in the synthetic data study for comparison}
    \label{fig:models}
\end{figure}

\textbf{Experimental Setting.} We evaluate the performance of the proposed algorithms with respect to the aforementioned state-of-the-art algorithms using synthetically generated datasets under various settings. We consider the \textit{Line, Grid and Erd\H{o}s-R\'enyi (with $5\%$ edge density)} structures as data generating models. A \textit{line graph} is a network where all vertices are arranged in a single continuous sequence. Each node has exactly two neighbors, except for the two endpoints which connect to only one neighbor each. A \textit{grid graph} is a structured lattice, where vertices are organized into rows and columns, and each node, except those on the boundaries, connects to its four immediate neighbors (up, down, left, right). Finally, an \textit{Erd\H{o}s-R\'enyi graph} is constructed with a fixed number of nodes, and each possible pair of nodes has an independent, identical probability ($5\%$ in our experiments) of being connected by an edge. Figure~\ref{fig:models} provides visual representations of these graph structures. For each structure, we consider six different simulation settings corresponding to the following $(n,p)$ combinations: $(100,20), (100,50), (100,100), (100,200), (500,100), (500,250)$.

{\color{black}For each fixed graph structure and $(n,p)$ combination, we generate 50 independent datasets with $6n/5$ observations following the data-generation framework in \cite{slawski2014estimationpositivedefinitemmatrices}.} First we construct the graph according to the given structure, and then assign weights to the edges by sampling randomly from $U(2,5)$ distribution to obtain the weighted adjacency matrix $A$. Defining $\delta = 1.05 \lambda_{max}(A)$, where $\lambda_{max}(A)$ is the maximum eigenvalue of $A$, we set $$\Omega^{*}=\delta I-A,\, \Omega_{0}=E \Omega^{*} E,$$ where $E$ is a diagonal matrix ensuring that the diagonal elements of $(\Omega_{0})^{-1}$ are $1$. We consider $\Omega_{0}$ to be our true precision matrix, and draw $n$ independent samples $\yb_{1},\yb_{2},\dots,\yb_{n}$ from $N(\mathbf{0},\Omega_{0}^{-1})$. {\color{black} The additional $n/5$ observations are used as test data to select slab inclusion probability \(q\) via the grid search described in Section~\ref{subsec:4.4:hyper}.}. We estimate the precision matrix using different algorithms as mentioned above, and evaluate the performance of the algorithms. Note again that SLTP only provides an estimate of the graph structure. 

\textbf{Evaluation Metrics.}
We first consider metrics for sparsity selection accuracy. Let $TP,\,FP,\,TN$ and $FN$ denote the true positives, false positives, true negatives and false negatives respectively. \textit{Matthews Correlation Coefficient (MCC)}~\citep{MATTHEWS1975442} and \textit{$F$-score}~\citep{Rijsbergen1979} of an estimated graph measure how well the true graph structure has been recovered. The $F$-score of a graph ranges between $0$ and $1$, with $F=1$ indicating a perfect structure recovery scenario, whereas MCC ranges between $-1$ and $1$, with $-1,0$ and $1$ denoting a perfectly inverse prediction, a random prediction and a perfect prediction respectively. {\color{black}We further report the \textit{Area Under the Receiver Operating Characteristic curve (AUROC)} as a measure of graph selection performance of the algorithms.} To assess the quality of estimation, we compute the \textit{Relative Estimation Error}, obtained as the Frobenius norm of the difference of the estimated and true precision matrix, and scaled by the Frobenius norm of the true precision matrix.

{\bf Results.} The MCC, $F$-score and the estimation errors corresponding to all the settings considered in our experiment are reported below in Table~\ref{tab:1:MCC},~\ref{tab:2:Fscore} and~\ref{tab:3:est_error} respectively. {\color{black}Results on AUROC have been included in Supplementary Section~\ref{App:B3:AUROC}.}

\begin{table}[!ht]
\centering
{
\setstretch{1.3}
\scriptsize
\caption{\textit{Matthews Correlation Coefficient (MCC)} for the benchmark and proposed algorithms across three data-generating models, namely, Line Graph, Grid Graph, and Erdős–Rényi Graph ($5\%$ edge density), and six \((n,p)\) configurations.}
\label{tab:1:MCC}
\begin{tabular}{c c c c c c c c}
\toprule
\textbf{Model} & $\textbf{(n,p)}$ & \textbf{GLASSO} & \textbf{CLIME} & \textbf{GGL} & \textbf{D-Gibbs} & \textbf{A-Gibbs} & \textbf{I-Gibbs} \\
\midrule
\multirow{6}{*}{Line Graph} & $(100,20)$ & $0.339362$ & $0.162940$ & $0.592964$ & $0.839393$ & $0.900792$ & $\textbf{0.910718}$ \\
 & $(100,50)$ & $0.344382$ & $0.366485$ & $0.492189$ & $0.803132$ & $0.866734$ & $\textbf{0.883758}$ \\
 & $(100,100)$ & $0.338153$ & $0.469339$ & $0.427420$ & $0.779140$  & $0.835501$ & $\textbf{0.856129}$\\
 & $(100,200)$ & $0.340614$ & $0.545414$ & $0.362629$ & $0.750760$ & $0.763749$ & $\textbf{0.806506}$ \\
 & $(500,100)$ & $0.324436$ & $0.442767$ & $0.391187$ & $0.889521$ & $0.929744$ & $\textbf{0.948483}$\\
 & $(500,250)$ & $0.328980$ & $0.472171$ & $0.310010$ & $0.918754$ & $0.950831$ & $\textbf{0.962887}$\\
\midrule
\multirow{6}{*}{Grid Graph} & $(100,20)$ & $0.418464$ & NaN & $0.589024$ & $0.762005$ & $0.782519$ & $\textbf{0.792389}$ \\
 & $(100,50)$ & $0.422413$ & $0.330798$ & $0.544509$ & $0.702119$ & $0.706322$ & $\textbf{0.718569}$\\
 & $(100,100)$ & $0.408616$ & $0.418420$ & $0.489635$ & $0.635538$ & $0.645766$ & $\textbf{0.648025}$ \\
 & $(100,200)$ & $0.384598$ & $0.499235$ & $0.462793$ & $0.624958$ & $0.576971$ & $\textbf{0.638008}$ \\
 & $(500,100)$ & $0.408196$ & $0.431054$ & $0.523006$ & $0.950699$ & $0.940638$ & $\textbf{0.963538}$ \\
 & $(500,250)$ & $0.396653$ & $0.587283$ & $0.469590$ & $0.937282$ & $0.915480$ & $\textbf{0.949102}$ \\
\midrule
\multirow{6}{*}{Erd\H{o}s-R\'enyi} & $(100,20)$ & $0.332911$ & $0.242688$ & $0.358103$ & $0.767684$ & $0.866433$ & $\textbf{0.878072}$ \\
& $(100,50)$ & $0.405567$ & $0.326772$ & $0.433457$ & $0.600928$ & $0.606377$ & $\textbf{0.614399}$\\
 & $(100,100)$ & $0.353774$ & $0.300434$ & $0.366720$ & $0.358896$ & $0.404450$ & $\textbf{0.407736}$\\
 & $(100,200)$ & $0.174604$ & $0.133601$ & $\textbf{0.194814}$ & $0.103091$ & $0.067407$ & $0.100008$ \\
 & $(500,100)$ & $0.443320$ & $0.432508$ & $0.484373$ & $0.805784$ & $0.833517$ & $\textbf{0.852969}$ \\
 & $(500,250)$ & $0.344685$ & $0.003510$ & $0.351726$ & $0.365776$ & $0.352156$ & $\textbf{0.369464}$ \\
\bottomrule
\end{tabular}
}
\end{table}

\begin{table}[!ht]
\centering
{\setstretch{1.3}
\scriptsize
\caption{\textit{$F$-scores} for the benchmark and proposed algorithms across three data-generating models, namely, Line Graph, Grid Graph, and Erdős–Rényi Graph ($5\%$ edge density), and six \((n,p)\) configurations.}
\label{tab:2:Fscore}
\begin{tabular}{c c c c c c c c}
\toprule
\textbf{Model} & $\textbf{(n,p)}$ & \textbf{GLASSO} & \textbf{CLIME} & \textbf{GGL} & \textbf{D-Gibbs} & \textbf{A-Gibbs} & \textbf{I-Gibbs} \\
\midrule
\multirow{6}{*}{Line Graph} & $(100,20)$ & $0.346516$ & $0.223613$ & $0.588181$ & $0.849258$ & $0.909195$ & $\textbf{0.918645}$\\
 & $(100,50)$ & $0.269845$ & $0.290592$ & $0.428643$ & $0.803568$ & $0.870196$ & $\textbf{0.887556}$ \\
 & $(100,100)$ & $0.235832$ & $0.384320$ & $0.332391$ & $0.777240$ & $0.837927$ & $\textbf{0.858570}$ \\
 & $(100,200)$ & $0.227745$ & $0.482143$ & $0.247804$ & $0.752525$ & $0.757187$ & $\textbf{0.807213}$ \\
 & $(500,100)$ & $0.222446$ & $0.350902$ & $0.290577$ & $0.885898$ & $0.928729$ & $\textbf{0.948192}$ \\ 
 & $(500,250)$ & $0.208240$ & $0.372885$ & $0.187323$ & $0.916166$ & $0.950052$ & $\textbf{0.962504}$ \\
\midrule
\multirow{6}{*}{Grid Graph} & $(100,20)$ & $0.480015$ & $0.299448$ & $0.626990$ & $0.799067$ & $0.814697$ & $\textbf{0.823955}$ \\
 & $(100,50)$ & $0.391348$ & $0.302863$ & $0.519985$ & $0.714203$ & $0.714059$ & $\textbf{0.724332}$ \\
 & $(100,100)$ & $0.346849$ & $0.370785$ & $0.435944$ & $0.628399$ & $0.650666$ & $\textbf{0.632516}$ \\
 & $(100,200)$ & $0.300669$ & $0.470017$ & $0.388500$ & $0.611755$ & $0.564937$ & $\textbf{0.614718}$ \\
 & $(500,100)$ & $0.332863$ & $0.355340$ & $0.461504$ & $0.952385$ & $0.942336$ & $\textbf{0.964802}$ \\
 & $(500,250)$ & $0.291539$ & $0.525477$ & $0.376971$ & $0.938167$ & $0.916677$ & $\textbf{0.949584}$ \\
\midrule
\multirow{6}{*}{Erd\H{o}s-R\'enyi} & $(100,20)$ & $0.280379$ & $0.190686$ & $0.290221$ & $0.756816$ & $0.864444$ & $\textbf{0.877599}$ \\
& $(100,50)$ & $0.376305$ & $0.291812$ & $0.394005$ & $0.601390$ & $0.601570$ & $\textbf{0.605846}$ \\
 & $(100,100)$ & $0.351698$ & $0.309478$ & $0.358247$ & $0.380263$ & $\textbf{0.420035}$ & $0.408998$ \\
 & $(100,200)$ & $0.207444$ & $0.172611$ & $\textbf{0.232161}$ & $0.054220$ & $0.035888$ & $0.050440$ \\
 & $(500,100)$ & $0.389738$ & $0.379392$ & $0.433820$ & $0.815546$ & $0.841990$ & $\textbf{0.859585}$ \\
 & $(500,250)$ & $0.346394$ & $0.026335$ & $0.351061$ & $0.343584$ & $0.342464$ & $\textbf{0.359817}$ \\
\bottomrule
\end{tabular}
}
\end{table}

Tables~\ref{tab:1:MCC} and~\ref{tab:2:Fscore} report the MCC and $F$-scores across all $18$ experimental settings. 
Across nearly all simulation settings, the proposed Gibbs samplers consistently achieve substantially higher MCC and $F$-scores than the competing methods. The sole exception, in which GGL outperforms the proposed algorithms, occurs under the Erd\H{o}s–R\'enyi model with $n=100$ and $p=200$ (and $245$ non-zero off-diagonal entries in the corresponding $\Omega_0$), a severely data-deprived regime in which all six algorithms attain notably low MCC and $F$-score values. In all remaining settings, the proposed algorithms demonstrate substantial improvements over the general GLASSO and CLIME methods, which do not incorporate the sign constraint; the magnitude of improvement ranges from approximately $100\%$ in most settings to as much as $200\%$ in several others. While GGL, which does account for the sign constraints, improves upon GLASSO and CLIME, the proposed samplers exhibit considerably superior performance relative to GGL across nearly all settings. 

Among the three proposed algorithms, all of which deliver significantly improved performance as described above, the interweaving I-Gibbs sampler achieves the highest overall performance. Specifically, I-Gibbs consistently attains the highest MCC and $F$-score values across all settings, reflecting a marked improvement in graph selection performance over existing state-of-the-art algorithms. The augmented A-Gibbs sampler yields results closely comparable to I-Gibbs, with differences typically appearing only at the third decimal place. In contrast, the graph selection performance of D-Gibbs is somewhat inferior to that of the other two Gibbs samplers, a finding that is consistent with its characterization as a local explorer, as discussed in Section~\ref{subsec:4.3:interweaving}.

Table~\ref{tab:3:est_error} presents the precision matrix estimation errors across the $18$ settings. Algorithms that incorporate the $M$-matrix constraint uniformly outperform GLASSO and CLIME, which do not account for this constraint. Compared to GGL, the proposed methods achieve lower estimation error in certain settings and remain broadly competitive across the remaining configurations. 
In summary, the three proposed Gibbs samplers yield substantial improvements in graph structure estimation under the $M$-matrix constraint. With respect to graph selection, I-Gibbs and A-Gibbs demonstrate the strongest performance, whereas D-Gibbs proves most effective for precision matrix estimation.

\begin{table}[!ht]
\centering
{\setstretch{1.3}
\scriptsize
\caption{\textit{Relative Estimation Errors} (in Frobenius norm) for the benchmark and proposed algorithms across three data-generating models, namely, Line Graph, Grid Graph, and Erdős–Rényi Graph ($5\%$ edge density), and six \((n,p)\) configurations.}
\label{tab:3:est_error}
\begin{tabular}{c c c c c c c c}
\toprule
\textbf{Model} & $\textbf{(n,p)}$ & \textbf{GLASSO} & \textbf{CLIME} & \textbf{GGL} & \textbf{D-Gibbs} & \textbf{A-Gibbs} & \textbf{I-Gibbs} \\
\midrule
\multirow{6}{*}{Line Graph} & $(100,20)$ & $0.402249$ & $0.580012$ & $0.193142$ & $\textbf{0.170640}$ & $0.172726$ & $0.171740$ \\
 & $(100,50)$ & $0.537159$ & $0.435991$ & $0.229762$ & $\textbf{0.191810}$ & $0.195054$ & $0.192894$ \\
 & $(100,100)$ & $0.577636$ & $0.416646$ & $0.265771$ & $0.221955$ & $0.223213$ & $\textbf{0.219658}$ \\
 & $(100,200)$ & $0.564507$ & $0.395472$ & $0.317985$ & $0.260906$ & $0.261461$ & $\textbf{0.257024}$ \\
 & $(500,100)$ & $0.389555$ & $0.130876$ & $0.104114$ & $0.076846$ & $0.078304$ & $\textbf{0.076456}$\\
 & $(500,250)$ & $0.388881$ & $0.160640$ & $0.122187$ & $0.082941$ & $0.084125$ & $\textbf{0.082209}$\\
\midrule
\multirow{6}{*}{Grid Graph} & $(100,20)$ & $0.368722$ & $0.538054$ & $\textbf{0.246963}$ & $0.260354$ & $0.265003$ & $0.260866$ \\
 & $(100,50)$ & $0.413582$ & $0.381189$ & $\textbf{0.279902}$ &  $0.309609$ & $0.315558$ & $0.312877$\\
 & $(100,100)$ & $0.438997$ & $0.372740$ & $\textbf{0.303163}$ & $0.346133$ & $0.354197$ & $0.348000$\\
 & $(100,200)$ & $0.464642$ & $0.411157$ & $\textbf{0.306747}$ & $0.364031$ & $0.386358$ & $0.364370$\\
 & $(500,100)$ & $0.259640$ & $0.152300$ & $0.126216$ & $\textbf{0.103256}$ & $0.108265$ & $0.103862$\\
 & $(500,250)$ & $0.281666$ & $0.200332$ & $0.130323$ & $\textbf{0.112434}$ & $0.122991$ & $0.112953$\\
\midrule
\multirow{6}{*}{Erd\H{o}s-R\'enyi} & $(100,20)$ & $0.617947$ & $0.647322$ & $0.236652$ & $0.178985$ & $0.178695$ & $0.178519$\\
& $(100,50)$ & $0.478910$ & $0.399931$ & $0.363733$ & $0.334067$ & $0.339354$ & $0.335502$ \\
 & $(100,100)$ & $0.412339$ & $0.372308$ & $\textbf{0.346040}$ & $0.471846$ & $0.486807$ & $0.474448$\\
 & $(100,200)$ & $0.381610$ & $0.358277$ & $\textbf{0.337663}$ & $0.422962$ & $0.385606$ & $0.417079$\\
 & $(500,100)$ & $0.242779$ & $0.179009$ & $0.177906$ & $0.161810$ & $0.166856$ & $0.163581$ \\
 & $(500,250)$ & $0.255073$ & $0.562434$ & $\textbf{0.235388}$ & $0.301159$ & $0.308389$ & $0.303388$ \\
\bottomrule
\end{tabular}
}
\end{table}

\subsection{Financial Time-Series Data} \label{subsec:5.2:DataAnalysis}
In this section, we analyze financial time-series data and compare the performance of the proposed algorithms with existing benchmarks. The $M-$matrix constraint is natural in this context, since latent market factors induce positive dependence among stock prices within related market sectors~\citep{Hennessy2002TheUO, MULLER2005434}. This provides a statistical justification for graphical model selection under $M$-matrix constraints; see, for example, \citet{wang2020learninghighdimensionalgaussiangraphical} and \citet{ying2023adaptiveestimationgraphicalmodels}.

We consider five industrial sectors classified according to the Global Industry Classification Standard (GICS): Information Technology, Financials, Health Care, Consumer Discretionary, and Consumer Staples. The list of S\&P 500 constituent companies, including ticker symbols and GICS sector classifications, was obtained from Wikipedia's 'List of S\&P 500 companies' page. The data consists of \(p=286\) stocks from these sectors included in the S\&P 500 index. We collect daily closing prices from January 2022 to December 2023, yielding a total of \(n=501\) observations per stock. We compute daily log-returns and rescale them by multiplying with \(100\). Specifically, if \(P_{i,j}\) denotes the closing price of the \(j\)-th stock on the \(i\)-th trading day, then the \((i,j)\)-th entry of the log-return matrix \(X\in\mathbb{R}^{501\times 286}\) is
\[
X_{i,j}
=
100\log\left(\frac{P_{i+1,j}}{P_{i,j}}\right),
\qquad
i=1,\ldots,501,\quad j=1,\ldots,286.
\]

We treat the rows of \(X\) as independent and identically distributed realizations of a random vector representing the scaled daily log-returns of the \(286\) stocks.

Estimating the covariance matrix of these log-returns is of fundamental importance in financial econometrics and quantitative portfolio management~\citep{fan2015overviewestimationlargecovariance,c080ac4da3944540a513f97f620bd1b0}. The covariance matrix captures the co-movement structure of asset returns and serves as a key input for tasks such as mean--variance portfolio optimization, risk forecasting, and asset allocation~\citep{e5a1bb8f-41b7-35c6-95cd-8b366d3e99bc}. 

Moreover, two stocks are conditionally uncorrelated given the others if and only if the corresponding precision matrix entry is zero~\citep{lauritzen1996graphical,glasso2009}. Thus, estimating a sparse precision matrix can be viewed as recovering a financial network that captures correlations between stocks after accounting for common market factors. Such networks provide insight into sectoral interactions, clustering behavior, and potential pathways of systemic risk and contagion~\citep{article1,article2}. The $M$-matrix constraint further enhances interpretability and stability by enforcing positive dependence, consistent with economic intuition, and often leads to improved performance in high-dimensional settings. 

Since the true underlying precision matrix is unknown, direct evaluation of the estimated graph through the true positive rate, false positive rate, or estimation error is not possible. However, stocks belonging to the same industrial sector are expected to exhibit stronger dependence. This motivates the use of \textit{modularity}~\citep{Newman_2006} as an external measure for assessing the quality of the estimated graph, with the GICS sector labels serving as the reference community structure. Modularity is a widely used measure to evaluate community detection in networks. For a graph $\mathcal{G}$ with vertex set $V$, edge set $E$ and the adjacency matrix $A$, the modularity of the graph is defined as
\begin{align*}
    Q := \frac{1}{2|E|} \displaystyle \sum_{\substack{i,j=1 \\ }}^{|V|}\left( A_{ij} -\frac{d_{i}d_{j}}{2 |E|} \right) I(c_{i}=c_{j})
\end{align*}
where, $d_{i}$ denotes the degree of the $i$-th node, $c_{i}$ is the type/community of the $i$-th node, and as defined in Section~\ref{sec:2:setup}, $I(a=b)=1$, when $a=b$, and otherwise it is $0$. The value of $Q$ for any graph or network ranges between $-0.5$ and $1$. In our context, a high value of modularity indicates more connections between the stocks belonging to the same sector and lower connection between stocks belonging to different sectors. Hence, we expect the modularity to be high for a well-estimated precision matrix. We compare the performance of the proposed D-Gibbs, A-Gibbs and I-Gibbs sampler with GLASSO~\citep{glasso2009}, CLIME~\citep{Cai01062011} and GGL~\citep{article}.  {\color{black}For the proposed samplers, the slab inclusion probability \(q\) is selected via the grid-search procedure described in Section~\ref{subsec:4.4:hyper}.
} For GLASSO, CLIME and GGL, the respective hyperparameters have been determined using 5-fold cross-validation. Here also, sparsity is induced by thresholding the off-diagonal entries at $10^{-3}$ in CLIME algorithm, following \cite{Cai01062011}. The values of modularity for various algorithms are provided in Table~\ref{tab:4:modularity}. 
{\color{black} Natural uncertainty quantification through the posterior is a key advantage of Bayesian methods. Following~\cite{Barbieri_2004}, we use marginal inclusion probabilities to assess the uncertainty associated with the graphs selected by the proposed algorithms. Further details and corresponding results are provided in Supplementary Section~\ref{app:D:DA_uncertainty}.}

\begin{table}[!ht]
    \centering
    \caption{Comparison of modularity scores for precision matrix estimation algorithms applied to S\&P 500 financial time-series data from five GICS sectors. The graph induced by each estimated precision matrix is evaluated using modularity \(Q\), with sector labels serving as the reference community structure.}
    \label{tab:4:modularity}
    \footnotesize
    \begin{tabular}{c c c c c c c} 
    \toprule 
    \textbf{Algorithm} & \textbf{GLASSO} & \textbf{CLIME} & \textbf{GGL} & \textbf{D-Gibbs} & \textbf{A-Gibbs} & \textbf{I-Gibbs} \\ 
    \midrule
    \textbf{Modularity ($Q$)} & 0.136365 & 0.032529 & 0.283264 & 0.151645 & 0.454814 & \textbf{0.570447}  \\ 
    \bottomrule
    \end{tabular}
\end{table}

\noindent
As reported in Table~\ref{tab:4:modularity}, both A-Gibbs and I-Gibbs samplers yield significantly higher modularity values, indicating more effective community detection than GLASSO, CLIME and GGL. The low modularity achieved by D-Gibbs can be attributed to its recovery of a considerably larger number of \textit{between-sector} edges, as shown in Table~\ref{tab:7:uncertainty}.
\begin{figure}[!ht]
    \centering
    \begin{subfigure}[b]{0.3\textwidth}
        \centering
        \includegraphics[width=0.82\textwidth]{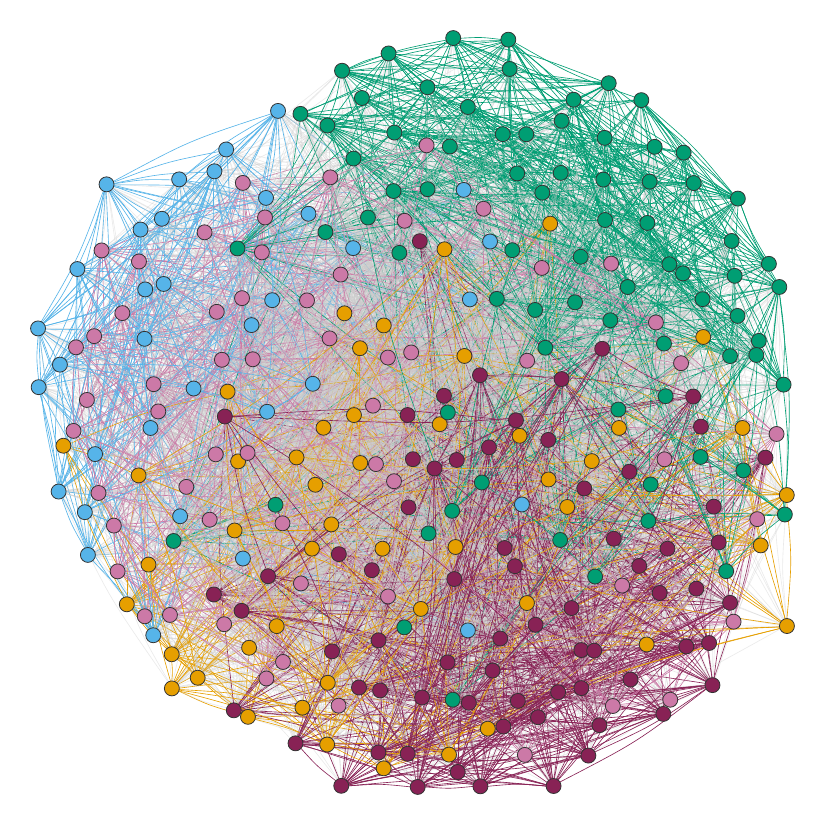}
        \caption{GLASSO}
    \end{subfigure}
    \hfill
    \begin{subfigure}[b]{0.3\textwidth}
        \centering
        \includegraphics[width=0.82\textwidth]{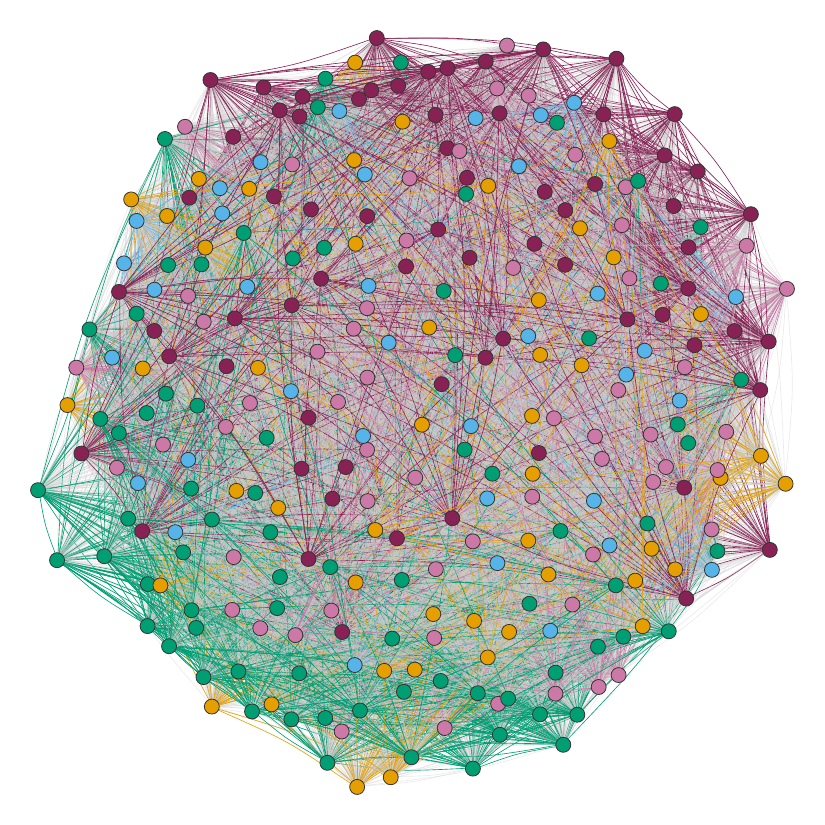}
        \caption{CLIME}
    \end{subfigure}
    \hfill
    \begin{subfigure}[b]{0.3\textwidth}
        \centering
        \includegraphics[width=0.82\textwidth]{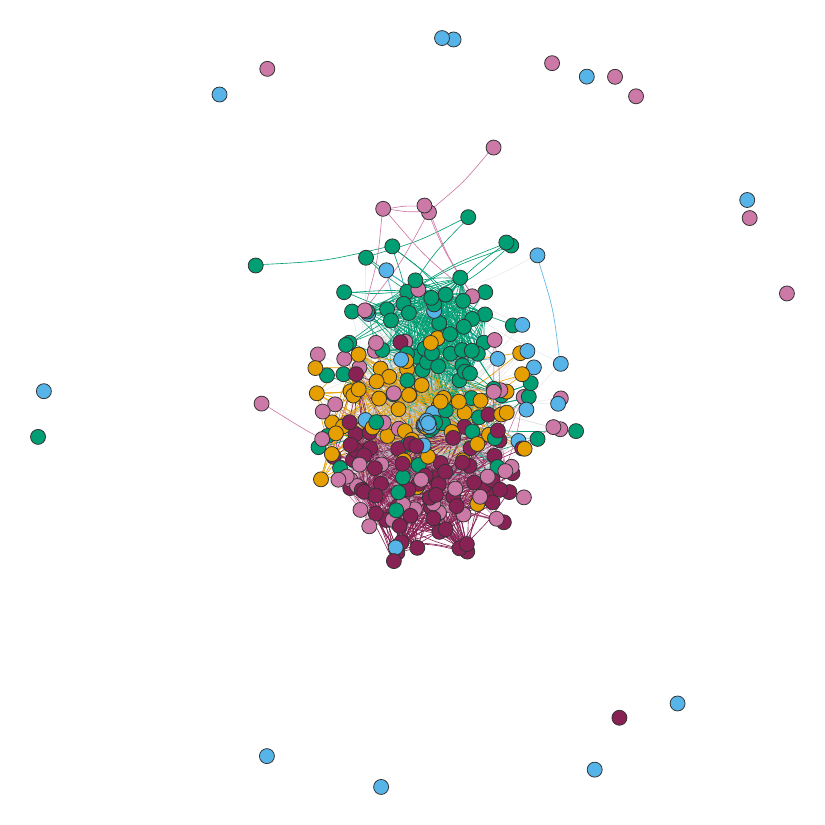}
        \caption{GGL}
    \end{subfigure}
    
    \vspace{0.3cm}  
    
    \begin{subfigure}[b]{0.3\textwidth}
        \centering
        \includegraphics[width=0.82\textwidth]{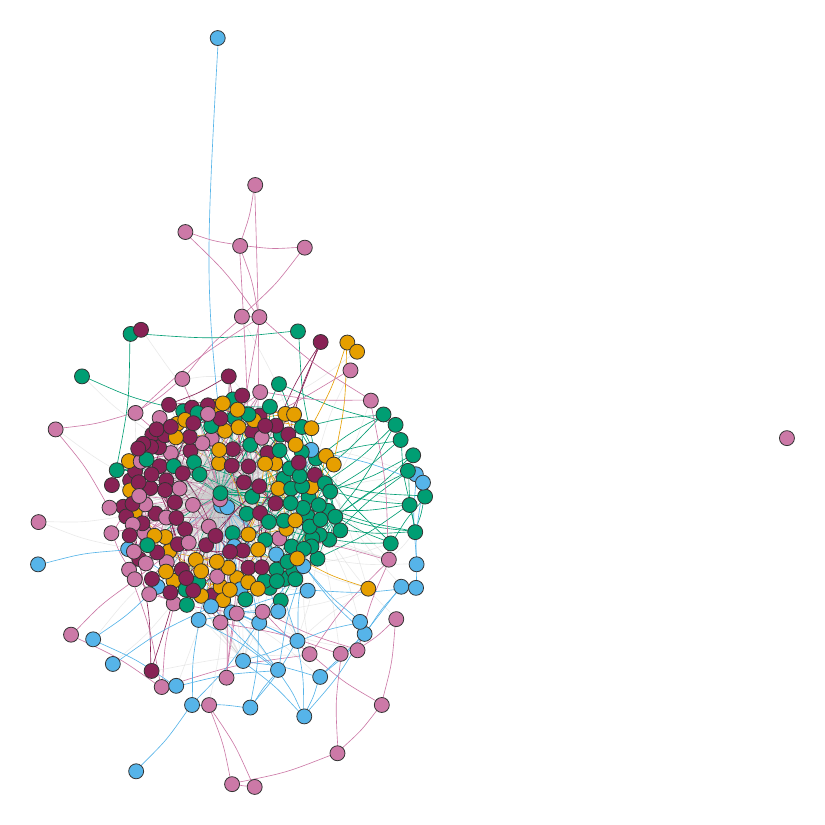}
        \caption{D-Gibbs}
    \end{subfigure}
    \hfill
    \begin{subfigure}[b]{0.3\textwidth}
        \centering
        \includegraphics[width=0.82\textwidth]{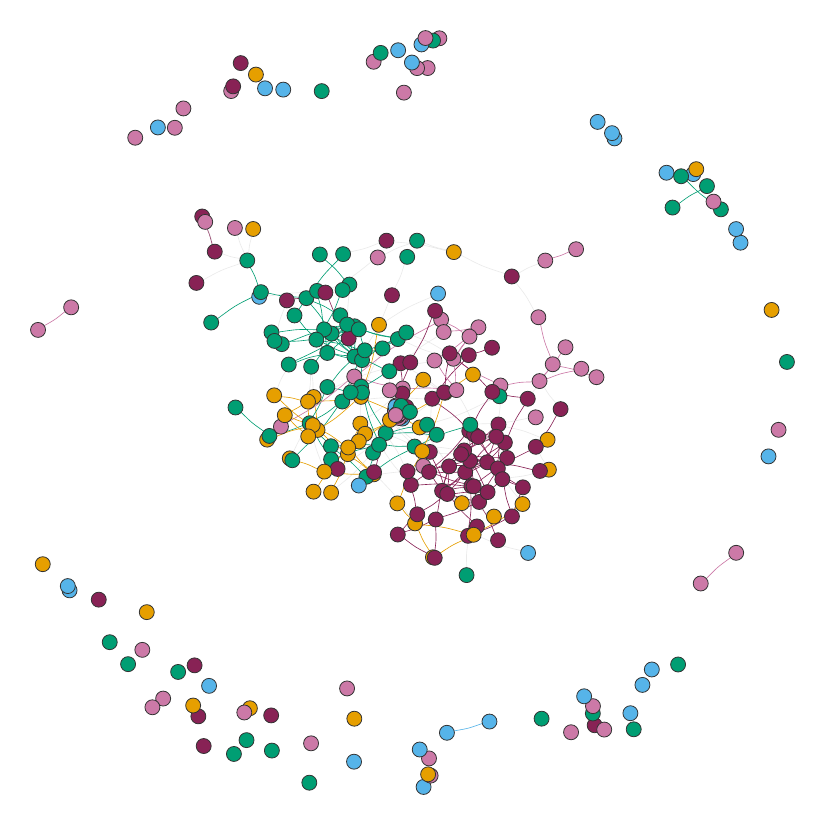}
        \caption{A-Gibbs}
    \end{subfigure}
    \hfill
    \begin{subfigure}[b]{0.3\textwidth}
        \centering
        \includegraphics[width=0.82\textwidth]{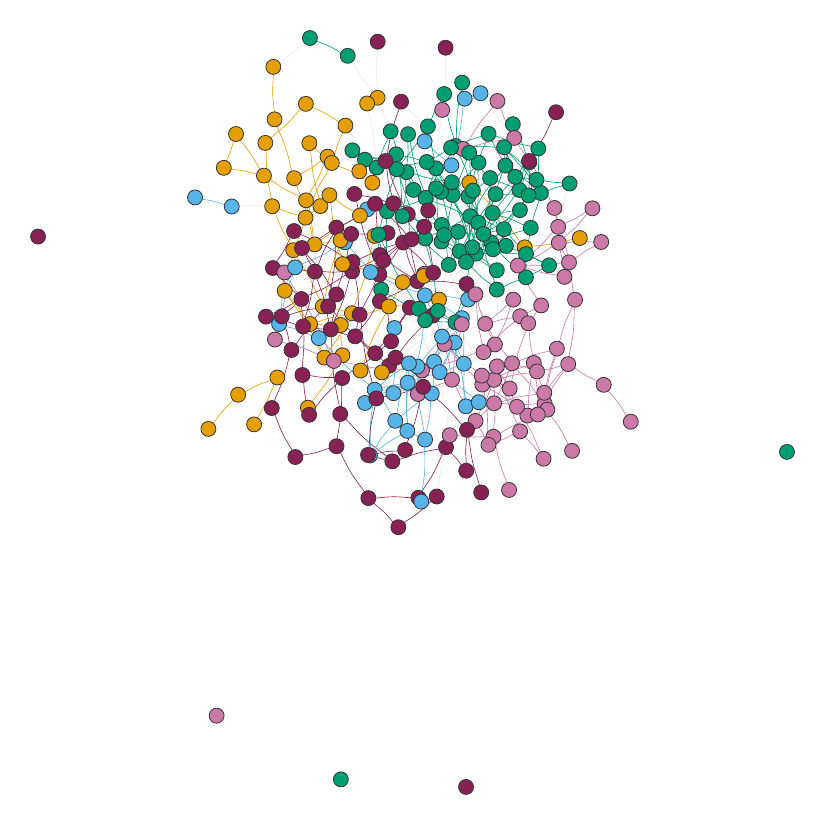}
        \caption{I-Gibbs}
    \end{subfigure}
    \caption{Network visualization of the estimated precision matrices for daily log-returns of 286 stocks from January 2022 to December 2023, using (a) GLASSO, (b) CLIME, (c) GGL, (d) D-Gibbs, (e) A-Gibbs and (f) I-Gibbs. The colored edges indicate edges that connect stocks from same sector, and the gray edges represent the cross-sector edges.}
    \label{fig:network_detection}
\end{figure}
Figure~\ref{fig:network_detection} depicts the graph structures estimated by the various algorithms under comparison. The figure highlights clear differences in the recovered network topologies across methods. In particular, the graphs obtained from the augmented and interweaving Gibbs samplers exhibit more coherent community structure, with stronger within-sector connectivity and fewer spurious cross-sector edges. In contrast, the graphs estimated by competing methods appear either overly dense or fail to capture the underlying modular organization effectively. This visual evidence is consistent with the quantitative results reported in Table~\ref{tab:4:modularity}, where the augmented and interweaving methods achieve substantially higher modularity scores, indicating improved recovery of meaningful structural patterns in the data.

\section{Discussion and Conclusion}\label{sec:conclusion}

\noindent
We proposed a scalable Bayesian framework for sparse precision matrix estimation under $M-$matrix constraints using a generalized posterior based on the convex D-trace loss. By avoiding the Gaussian likelihood and its log-determinant term, the method remains computationally tractable in high dimensions while supporting sparsity learning and uncertainty quantification through spike-and-slab priors. 

Our main methodological contribution is the development of efficient posterior samplers. A novel matrix-variate augmentation induces conditional independence among precision matrix entries, enabling joint updates and fast matrix-normal sampling that substantially reduces computational cost. An interweaving sampler further improves mixing by combining augmented and direct parameterizations. Simulations and real-data analyses show improved graph recovery and competitive precision matrix estimation relative to existing methods while maintaining scalability. In financial data, the estimated networks recover meaningful sectoral structure, highlighting the practical utility of the approach. 

Future work includes studying posterior contraction and graph selection consistency under the generalized framework, as well as extending the methodology to dynamic graphical models. More broadly, the combination of generalized Bayesian inference and augmentation-based computation may be useful for large-scale structured covariance estimation problems.

\section{Disclosure statement}\label{disclosure::statement}
The authors do not have any conflicts of interests to declare. 

\section{Data Availability Statement}
The financial data used in this study consist of daily adjusted closing prices for S\&P 500 constituent stocks from January 2022 to December 2023, obtained from Yahoo Finance via the \texttt{quantmod} R package (version 0.4.29). Sector classifications were obtained from the Wikipedia page \url{https://en.wikipedia.org/wiki/List_of_S%26P_500_companies}.

\section*{Declaration of AI-Assisted Technologies}

\noindent During the preparation of this manuscript, the authors used ChatGPT (GPT-6, OpenAI) for language refinement and coding assistance. All AI-assisted content was reviewed and verified by the authors, who take full responsibility for the final manuscript.

\begin{center}

{\large\bf SUPPLEMENTARY MATERIAL}

\end{center}

\noindent
The Supplementary Document contains proofs of various lemmas and additional experimental details. An anonymous version of the source code is available at \url{https://github.com/srnlrh/DAI-Gibbs}.
\bibliography{bibliography.bib}

\newpage
\clearpage
\setcounter{page}{1}
\phantomsection\label{supplementary-material}
\bigskip
\setcounter{equation}{0}
\renewcommand{\theequation}{S\arabic{equation}}

\setcounter{figure}{0}
\renewcommand{\thefigure}{S\arabic{figure}}

\setcounter{table}{0}
\renewcommand{\thetable}{S\arabic{table}}

\setcounter{algorithm}{0}
\renewcommand{\thealgorithm}{S\arabic{algorithm}}
\setcounter{lemma}{0}
\renewcommand{\thelemma}{S\arabic{lemma}}
\begin{center}

{\large\bf Supplementary Document for "Bayesian Graphical Models under Positivity Constraints: A Scalable Generalized Likelihood Approach"}

\end{center}
\setcounter{section}{0}
\renewcommand{\thesection}{\Alph{section}}

\setcounter{subsection}{0}
\renewcommand{\thesubsection}{\thesection.\arabic{subsection}}
  
\section{Proof of the Lemmas} \label{app:pf:lemma2}

\subsection{Proof of Lemma~\ref{lemma:2:post:proper}} \label{app:A1:pf:l2}

Let
\[
\xi=(\omega_{12},\omega_{13},\ldots,\omega_{1p},\omega_{23},\ldots,\omega_{p-1,p})^{T}
\]
denote the vector of off-diagonal entries of $\Omega$, and let
\[
\delta=(\omega_{11},\ldots,\omega_{pp})^{T}
\]
denote the vector of diagonal entries. Let $\mathcal L=\{0,1\}^{\binom p2}$ be the collection of all sparsity patterns for $\xi$. For $l\in\mathcal L$, let
$d_l$= number of active off-diagonal entries under the pattern $l$, and let
$\mathcal M_l$ denote the corresponding subspace of off-diagonal vectors satisfying that pattern.

Using the spike-and-slab representation, the normalizing constant associated with the posterior kernel in Lemma~\ref{lemma:1:post:jt} can be written as
\[
Z
=
\sum_{l\in\mathcal L}
q^{d_l}(1-q)^{\binom p2-d_l}
\left(\frac{2}{\tau\sqrt{2\pi}}\right)^{d_l}
Z_l,
\]
where
\[
Z_l
=
\int_{\mathcal S_l^*(p)}
\exp\left\{
n\operatorname{tr}(\Omega)
-\frac n2\operatorname{tr}(\Omega^2S)
-\frac{1}{2\tau^2}\|\xi_l\|_2^2
\right\}
\,d\delta\,d\xi_l .
\]
Here $\xi_l$ denotes the subvector of active off-diagonal entries under the pattern $l$, and
$\mathcal S_l^*(p)$ denotes the corresponding section of $\mathcal S^*(p)$ for which
$\xi\in\mathcal M_l$.

It is enough to show that $Z_l<\infty$ for every $l\in\mathcal L$, since $\mathcal L$ is finite. Fix $l\in\mathcal L$, and write
\[
\theta_l=(\delta^{T},\xi_l^{T})^{T}\in\mathbb R^{p+d_l}.
\]
Since $\Omega\mapsto \operatorname{tr}(\Omega^2S)$ is a quadratic form in the free parameters, there exists a nonnegative definite matrix $Q_l(S)$ such that
\[
\operatorname{tr}(\Omega^2S)=\theta_l^{T}Q_l(S)\theta_l .
\]
Define
\[
H_l
=
nQ_l(S)
+
\begin{pmatrix}
0_{p\times p} & 0\\
0 & \tau^{-2}I_{d_l}
\end{pmatrix}.
\]
Then
\[
n\operatorname{tr}(\Omega)
-\frac n2\operatorname{tr}(\Omega^2S)
-\frac{1}{2\tau^2}\|\xi_l\|_2^2
=
n\mathbf 1_p^{T}\delta
-\frac12\theta_l^{T}H_l\theta_l .
\]

We next show that \(H_l\) is positive definite on the joint parameter vector
\(\theta_l=(\delta^{T},\xi_l^{T})^{T}\). For any \(\theta_l\), let
\(\Omega=\Omega(\theta_l)\) denote the corresponding symmetric matrix under the sparsity pattern \(l\). Then
\[
\theta_l^{T}H_l\theta_l
=
n\operatorname{tr}(\Omega^2S)
+
\tau^{-2}\|\xi_l\|_2^2
=
n\operatorname{tr}(\Omega S\Omega)
+
\tau^{-2}\|\xi_l\|_2^2 .
\]
Since \(S\) is nonnegative definite,
\[
\operatorname{tr}(\Omega S\Omega)
=
\|S^{1/2}\Omega\|_F^2
\ge 0.
\]
Therefore \(H_l\) is nonnegative definite. To prove positive definiteness, suppose that
\[
\theta_l^{T}H_l\theta_l=0.
\]
Then both nonnegative terms above must be zero. In particular,
\[
\|\xi_l\|_2^2=0,
\]
and hence \(\xi_l=0\). Since all inactive off-diagonal entries are already fixed at zero by the sparsity pattern \(l\), it follows that all off-diagonal entries of \(\Omega\) are zero. Thus
\[
\Omega=\operatorname{diag}(\delta).
\]
Consequently,
\[
0=\operatorname{tr}(\Omega S\Omega)
=
\sum_{j=1}^{p}S_{jj}\omega_{jj}^2 .
\]
Since \(S_{jj}>0\) for every \(j=1,\ldots,p\), we obtain
\[
\omega_{11}=\cdots=\omega_{pp}=0.
\]
Hence \(\delta=0\), and therefore \(\theta_l=0\). This proves that
\[
\theta_l^{T}H_l\theta_l>0
\qquad
\text{for every } \theta_l\neq 0,
\]
so \(H_l\) is positive definite.
Consequently,
\[
Z_l
\le
\int_{\mathbb R^{p+d_l}}
\exp\left\{
n\mathbf 1_p^{T}\delta
-\frac12\theta_l^{T}H_l\theta_l
\right\}
\,d\theta_l .
\]
Since $H_l$ is positive definite, the last integral is a finite Gaussian integral. More explicitly,
\[
\int_{\mathbb R^{p+d_l}}
\exp\left\{
a_l^{T}\theta_l
-\frac12\theta_l^{T}H_l\theta_l
\right\}
\,d\theta_l
=
(2\pi)^{(p+d_l)/2}|H_l|^{-1/2}
\exp\left\{\frac12 a_l^{T}H_l^{-1}a_l\right\}
<\infty ,
\]
where
\[
a_l=(n\mathbf 1_p^{T},0_{d_l}^{T})^{T}.
\]
Therefore $Z_l<\infty$ for every $l\in\mathcal L$. Since the number of sparsity patterns is finite,
\[
Z
=
\sum_{l\in\mathcal L}
q^{d_l}(1-q)^{\binom p2-d_l}
\left(\frac{2}{\tau\sqrt{2\pi}}\right)^{d_l}
Z_l
<\infty .
\]
Hence the joint generalized posterior distribution of $\Omega$, given $\Yb$, is proper.

\subsection{Proof of Lemma~\ref{lemma:3:post:comp}} \label{app:A2:pf:l3}
We write
\begin{align*}
\operatorname{tr}(\Omega^2 S)
=
\sum_{i=1}^{p}
\left[
s_{ii}\omega_{ii}^{2}
+2\omega_{ii}\sum_{j\neq i}\omega_{ij}s_{ij}
\right]
+\sum_{1\leq i<j\leq p}(s_{ii}+s_{jj})\omega_{ij}^{2}
\quad
+2\sum_{i=1}^{p}
\sum_{\substack{1\leq j<k\leq p\\ j\neq i,\;k\neq i}}
\omega_{ij}\omega_{ik}s_{jk}.
\end{align*}
Using this and the expression of $\pi_{D}(\Omega|\Yb)$ in equation~\ref{eqn:4:post:exact}, we obtain the following conditional posterior distributions.
\begin{align*}
\pi(\omega_{ij} \mid \Omega_{-(ij)}, \Yb) &\propto \exp \left\{ -\frac{n}{2} (a_{ij} \omega_{ij}^2 + 2b_{ij} \omega_{ij}) \right\} \left( I_{\{0\}}(\omega_{ij}) + \frac{2q}{(1-q)\sqrt{2\pi\tau^2}} I_{(-\infty,0)}(\omega_{ij}) \right) \\
&= I_{\{0\}}(\omega_{ij}) + c_{ij} \frac{\sqrt{n a_{ij}}}{\sqrt{2\pi}} \exp \left\{ -\frac{n a_{ij}}{2} \left( \omega_{ij} + \frac{b_{ij}}{a_{ij}} \right)^2 \right\} I_{(-\infty,0)}(\omega_{ij})
\end{align*}
with $a_{ij},b_{ij}$ and $c_{ij}$ as defined earlier in equation~\ref{eqn:5:fullcond:offdiag}. Writing $q_{ij} = \frac{c_{ij}}{1+c_{ij}}$, we obtain the conditional posterior distribution in equation~\ref{eqn:5:fullcond:offdiag} for $1 \leq i<j \leq p$. 

Similarly, the conditional posterior distributions for the diagonal entries of $\Omega$ are computed as follows. For $i=1,2,\dots,p$,
\begin{align*}
\pi(\omega_{ii} \mid \Omega_{-(ii)}, \Yb) &\propto \exp \left( n \omega_{ii} - \frac{n}{2} s_{ii} \omega_{ii}^2 - n d_i \omega_{ii} \right) I_{(0,\infty)}(\omega_{ii}) \\
&\propto  \exp \left[ -\frac{n s_{ii}}{2} \left( \omega_{ii} - \frac{1 - d_i}{s_{ii}} \right)^2 \right] I_{(0,\infty)}(\omega_{ii})
\end{align*}
with $d_{i} = \displaystyle \sum_{i'\neq i} \omega_{ii'}s_{ii'}$. Therefore, the posteriors of the diagonal entries of $\Omega$ are as in equation~\ref{eqn:6:fullcond:diag}.

\subsection{Proof of Lemma~\ref{lemma:4:compDirect}} \label{app:A3:pf:l4}
First, note that step 3 of algorithm~\ref{algo:1:directgibbs} has computational complexity $\mathcal{O}(np^{2})$. However, this step is performed only once in the beginning of the algorithm. Steps 5-22 are repeated in each of the $T$ iterations of the Gibbs sampler. 

Among these, steps 8-16 sample the off-diagonal entries of $\Omega$ and require $\mathcal{O}(p)$ computations for the vector multiplication in computing $b$ in step 8, and $\mathcal{O}(1)$ computation for all the other steps. These steps with complexity $\mathcal{O}(p)$ are performed for each of the $\binom{p}{2}$ distinct off-diagonal entries. 

Finally, steps 17-18 also require $\mathcal{O}(p)$ computations, and are repeated $\mathcal{O}(p)$ times, constituting $\mathcal{O}(p^2)$ complexity. Therefore, in total, the D-Gibbs algorithm described in~\ref{algo:1:directgibbs} has a computational complexity $\mathcal{O}(p^{3})$.  \hfill $\square$

\subsection{Proof of Lemma~\ref{lemma:5:augcond}} \label{app:A4:pf:l5}
The augmented posterior distribution of $\Omega$, conditioned on the augmentation variable $R$ and the observation matrix $\Yb$ can be written as
\begin{equation} \label{eqn:1:pf:lemma5}
    \pi(\Omega \mid R, \Yb) \propto \pi(R \mid \Omega, \Yb) L_{D}(\Omega \mid \Yb) \pi(\Omega)
\end{equation}
where,
\begin{align} \label{eqn:2:pf:lemma5}
    \pi(R \mid \Omega, \Yb) &\propto \exp \left( -\frac{1}{2} \operatorname{tr} \left[ (kI - nS)^{-1} (R - \Omega(kI - nS))^{T} (R - \Omega(kI - nS)) \right] \right) \nonumber \\
    & \propto \exp \left( -\frac{k}{2} \operatorname{tr}(\Omega^2) + \frac{n}{2} \operatorname{tr}(\Omega^2 S) + \operatorname{tr}(\Omega R) \right)
\end{align}

Plugging $\pi(R \mid \Omega, \Yb)$ in~\ref{eqn:1:pf:lemma5} in the expression of $\pi(\Omega \mid R, \Yb)$ in~\ref{eqn:2:pf:lemma5}, we obtain 
\begin{align*}
    \pi(\Omega \mid R, \Yb) &\propto \exp \left( -\frac{k}{2} \operatorname{tr}(\Omega^2) + \operatorname{tr}(\Omega R) + n \operatorname{tr}(\Omega) \right)  \prod_{i<j} \left( (1-q) I_{\{0\}}(\omega_{ij}) + \frac{2q}{\tau \sqrt{2\pi}} e^{-\frac{\omega_{ij}^2}{2\tau^2}} I_{(-\infty,0)}(\omega_{ij}) \right) \\
    &= \exp \left( -\frac{1}{2} \left[ \sum_{i<j} 2k \omega_{ij}^2 - 2 \sum_{i<j} \omega_{ij} (R_{ij} + R_{ji}) \right] \right)  \exp \left( n \sum_{i=1}^p \omega_{ii} - \frac{k}{2} \sum_{i=1}^p \omega_{ii}^2 + \sum_{i=1}^p \omega_{ii} R_{ii} \right) \\
    &\qquad \times \prod_{i<j} \left( (1-q) I_{\{0\}}(\omega_{ij}) + \frac{2q}{\tau \sqrt{2\pi}} e^{-\frac{\omega_{ij}^2}{2\tau^2}}I_{(-\infty,0)}(\omega_{ij}) \right)
\end{align*}

For $1 \leq i<j \leq p$, the posterior distribution of $\omega_{ij}$, conditioned on $R$ and $\Yb$ can be computed as follows.
\begin{align*}
& \pi(\omega_{ij} \mid R, \Yb, q, \tau^2) \\
\propto & \, e^{-\frac{1}{2} \left[ 2k \omega_{ij}^2 - 2(R_{ij} + R_{ji}) \omega_{ij} \right]} \left[ (1-q)I_{\{0\}}(\omega_{ij}) + q \frac{2}{\tau\sqrt{2\pi}} e^{-\frac{\omega_{ij}^2}{2\tau^2}} I_{(-\infty,0)}(\omega_{ij}) \right] \\
\propto & \, \exp \left[ -\frac{1}{2} \left\{ \left(2k + \frac{1}{\tau^2}\right) \omega_{ij}^2 - 2 (R_{ij} + R_{ji}) \omega_{ij} \right\} \right] \left[ I_{\{0\}}(\omega_{ij}) + \frac{2q}{(1-q)} \frac{1}{\tau\sqrt{2\pi}}I_{(-\infty,0)}(\omega_{ij}) \right] \\
= & \, \exp \left[ -\frac{k^*}{2} \left( \omega_{ij}^2 - 2 \frac{R_{ij} + R_{ji}}{k^*} \omega_{ij} \right) \right] \left[ I_{\{0\}}(\omega_{ij}) + \frac{2q}{(1-q)} \frac{1}{\tau\sqrt{2\pi}}I_{(-\infty,0)}(\omega_{ij}) \right]; \quad  k^* = 2k + \frac{1}{\tau^2} \\
= & \,I_{\{0\}}(\omega_{ij}) + c_{ij} \cdot \frac{\sqrt{k^*}}{\sqrt{2\pi}} \exp \left[ -\frac{k^*}{2} \left( \omega_{ij} - \frac{R_{ij} + R_{ji}}{k^*} \right)^2 \right] I_{(-\infty,0)}(\omega_{ij})
\end{align*}

Writing $c_{ij} = \frac{2q}{(1-q)} \cdot \frac{1}{\tau\sqrt{k^*}} \cdot \exp\left[ \frac{(R_{ij} + R_{ji})^2}{2k^*} \right] \Phi \left( -\frac{R_{ij}+R_{ji}}{\sqrt{k^{*}}}\right)$, and $q_{ij}^* = \frac{c_{ij}}{1 + c_{ij}}$, we have
\[
\pi(\omega_{ij} \mid R, \Yb, \tau^2, q) \propto (1 - q_{ij}^*) \delta_{0} + q_{ij}^* \mathcal{N}\left( \frac{R_{ij} + R_{ji}}{k^*}, \frac{1}{k^*} ;\, -\infty,0\right).
\]

Also, for $i=1,2,\dots,p,$
\begin{align*}
    \pi(\omega_{ii} \mid R_{ii}, \Yb) &\propto \exp \left[ (R_{ii} + n) \omega_{ii} - \frac{k}{2} \omega_{ii}^2 \right] I_{(0,\infty)}(\omega_{ii})\\
    &\propto \exp \left[ -\frac{k}{2} \left( \omega_{ii} - \frac{R_{ii} + n}{k} \right)^2 \right]I_{(0,\infty)}(\omega_{ii})
\end{align*}

Therefore, the conditional posterior distribution of the diagonal entries of $\Omega$, for $i=1,2,\dots,p$ are given by
\begin{align*}
    \omega_{ii} \mid R_{ii}, \Yb &\overset{\text{ind}}{\sim} \mathcal{N} \left( \frac{R_{ii} + n}{k}, \frac{1}{k}; 0,\infty \right)
\end{align*}
Hence the proof. \hfill $\square$

\subsection{Proof of Lemma~\ref{lemma:6:struct:gauss}} \label{app:A5:pf:l6}
The moment generating function (mgf) of $X$ is given by ---
\begin{align*}
M_X(P)
&= E \left[ \exp\left\{\operatorname{tr}(P^{T} X)\right\} \right] \\
&= E \left[ \exp\left\{\sqrt{k} \operatorname{tr} \left( P^{T} \{ V_1(I_p - UU^{T}) + V_2 A^{T} U^{T} \} \right)\right\} \right] \\
&= E \left[
\exp\left\{\sqrt{k} \operatorname{tr} \left( (I_p - UU^{T}) P^{T} V_1 \right)\right\}
\exp\left\{\sqrt{k} \operatorname{tr} \left( A^{T} U^{T} P^{T} V_2 \right)\right\}
\right]\\
&= M_{V_1}(\sqrt{k} P (I_p - UU^{T})) M_{V_2}(\sqrt{k} PUA) , \text{ since } V_1, V_2 \text{ are independent} \\
&= \exp \left[{\frac{k}{2} \operatorname{tr} \left( (I_p - UU^{T}) P^{T} P (I_p - UU^{T}) \right)} \right] \exp \left[ {\frac{k}{2} \operatorname{tr} \left( A^{T} U^{T} P^{T} PUA \right)} \right] \\
&= \exp \left[{\frac{k}{2} \operatorname{tr} \left( (I_p - UU^{T}) P^{T} P\right)} \right] , \text{ using } AA^{T} = I_n - U^{T}U, \\
&= \exp \left[{\frac{1}{2} \operatorname{tr} \left( (kI_p - nS) P^{T} P \right)} \right]
\end{align*}
Using the uniqueness of the MGF, $X \sim \mathcal{MN}_{p \times p} (0, I_p, kI_p - nS)$. 

After precomputing $A$, each draw of $X$ requires 
 $\mathcal{O}(p^{2})$  and $\mathcal{O}(np)$ flops to compute
samples $V_{1}$ and $V_{2}$ {\color{black}respectively}, and computing $V_{1}(I_{p}-UU^{T})+V_{2}A^{T}U^{T} $ requires an additional 
{\color{black}$\mathcal{O}(np^{2})$ flops}
Therefore, one round of sampling $X$ has a computational complexity of 
$\mathcal{O}(np^{2})$, when $p>n$.  \hfill $\square$

\section{Additional Experimental Results}
\label{app:A6:exp:AUCROC}

\subsection{Proportion of Positive Definite Samples in Proposed Gibbs Samplers}\label{sec:sup:PD}
As discussed in Section~\ref{sec:3:model}, the proposed Gibbs samplers do not guarantee that either posterior draws or point estimates are positive definite. This relaxation, however, yields substantial computational gains by enabling efficient posterior sampling through Gibbs updates. In Table~\ref{tab:5:PD_prop}, we report the proportion of positive definite posterior draws under the simulation settings considered in Section~\ref{subsec:5.1:simulation}. {\color{black}For the purely augmented sampler (A-Gibbs) and the interweaving sampler (I-Gibbs), a substantial fraction of posterior draws are positive definite in almost all settings. In contrast, the corresponding proportions for the direct Gibbs sampler (D-Gibbs) vary significantly- sometimes they are high, and sometimes they are substantially lower.} These findings suggest that augmentation not only improves sparsity selection, but also increases the likelihood that posterior draws satisfy positive definiteness, a desirable structural property of precision matrices.

\begin{table}[!ht]
\centering
{\setstretch{1.3}
\footnotesize
\caption{\textit{Proportion of Positive Definite samples} out of the $3000$ samples for the proposed Gibbs samplers. The proportion is reported for three data generating models, namely, Line graph, Grid graph and Erd\H{o}s-R\'enyi Graph (with $5\%$ edge-density), and six \((n,p)\) configurations.}
\label{tab:5:PD_prop}
\begin{tabular}{c c c c c}
\toprule
\textbf{Model} & $(n,p)$ & \textbf{D-Gibbs} & \textbf{A-Gibbs} & \textbf{I-Gibbs} \\
\midrule
\multirow{6}{*}{Grid Graph} & $(100,20)$ & 0.977773 & 0.987680 & 0.969973 \\
 & $(100,50)$ &  0.968927 & 0.990193 & 0.989800\\
 & $(100,100)$ & 0.944287 & 0.728107 & 0.985040 \\
 & $(100,200)$ & 0.806647 & 0.924553 & 0.943500 \\
 & $(500,100)$ & 1 & 0.999980 & 1 \\
 & $(500,250)$ & 1 & 1 & 1 \\
\midrule
\multirow{6}{*}{Erd\H{o}s-R\'enyi} & $(100,20)$ & 0.981633 & 0.984380 & 0.988627 \\
& $(100,50)$ & 0.967980 & 0.992453 & 0.984093 \\
 & $(100,100)$ & 0.544607 & 0.534313 & 0.949600 \\
 & $(100,200)$ & 1 & 1 & 1 \\
 & $(500,100)$ & 0.999100 & 0.998467 & 0.999980 \\
 & $(500,250)$ &  1 & 1 & 1  \\
\midrule
\multirow{6}{*}{Line Graph} & $(100,20)$ & 0.928573 & 0.943060 & 0.955793 \\
 & $(100,50)$ & 0.607133 & 0.681673 & 0.738800 \\
 & $(100,100)$ & 0.157527 & 0.508880 & 0.405813 \\
 & $(100,200)$ & 0.130827 & 0.864593 & 0.373833 \\
 & $(500,100)$ & 1 & 1 & 1\\
 & $(500,250)$ &  1 & 1 & 1 \\
\bottomrule
\end{tabular}
}
\end{table}

\subsection{Comparison of Execution Time of D-Gibbs, A-Gibbs and I-Gibbs}

\begin{figure}[h]
    \centering
    \begin{subfigure}[b]{0.32\textwidth}
        \centering
        \includegraphics[width=0.9\textwidth, height = 0.2\textheight]{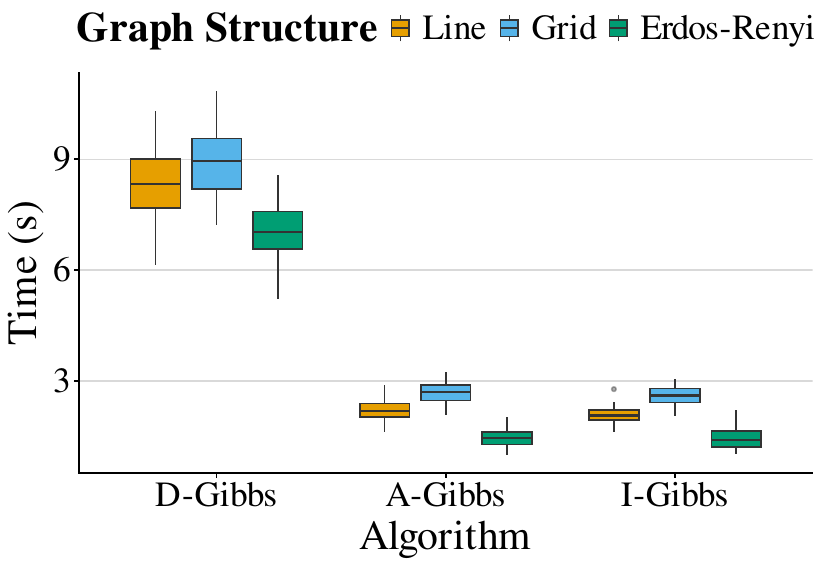}
        \caption{$(n,p)=(100,20)$}
    \end{subfigure}
    \begin{subfigure}[b]{0.32\textwidth}
        \centering
        \includegraphics[width=0.9\textwidth, height = 0.2\textheight]{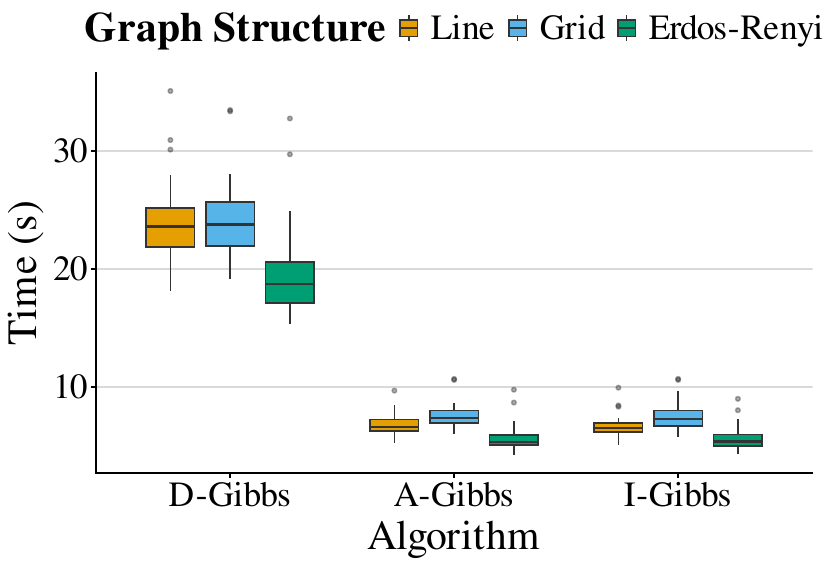}
        \caption{$(n,p)=(100,50)$}
    \end{subfigure}
    \begin{subfigure}[b]{0.32\textwidth}
        \centering
        \includegraphics[width=0.9\textwidth, height = 0.2\textheight]{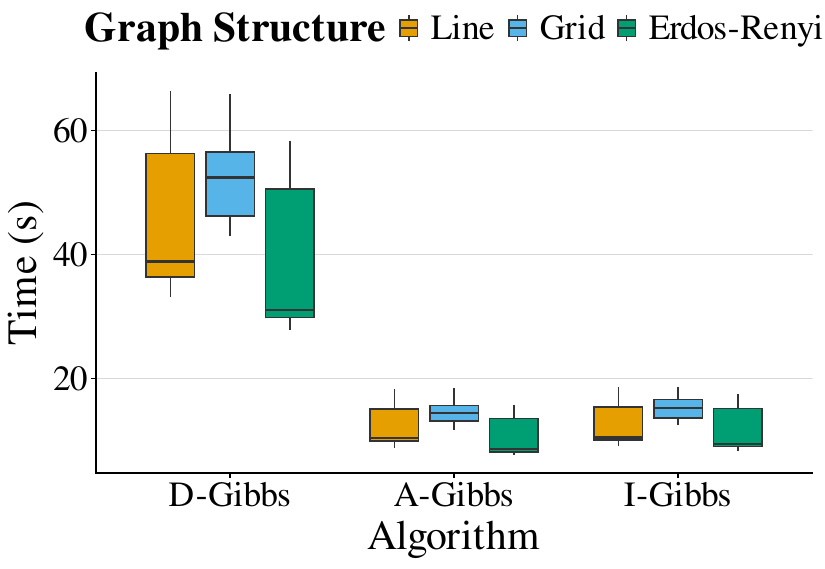}
        \caption{$(n,p)=(100,100)$}
    \end{subfigure}
    
    \vspace{0.3cm}  
    
    \begin{subfigure}[b]{0.32\textwidth}
        \centering
        \includegraphics[width=\textwidth, height = 0.12\textheight]{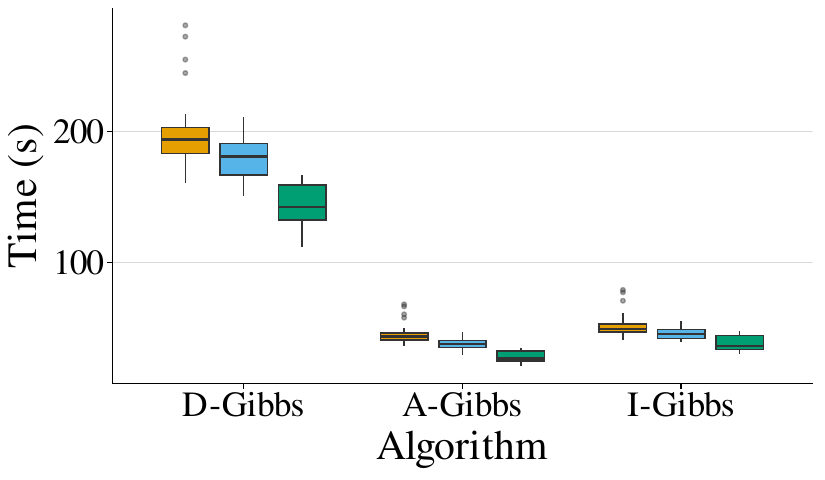}
        \caption{$(n,p)=(100,200)$}
    \end{subfigure}
    \begin{subfigure}[b]{0.32\textwidth}
        \centering
        \includegraphics[width=\textwidth, height = 0.12\textheight]{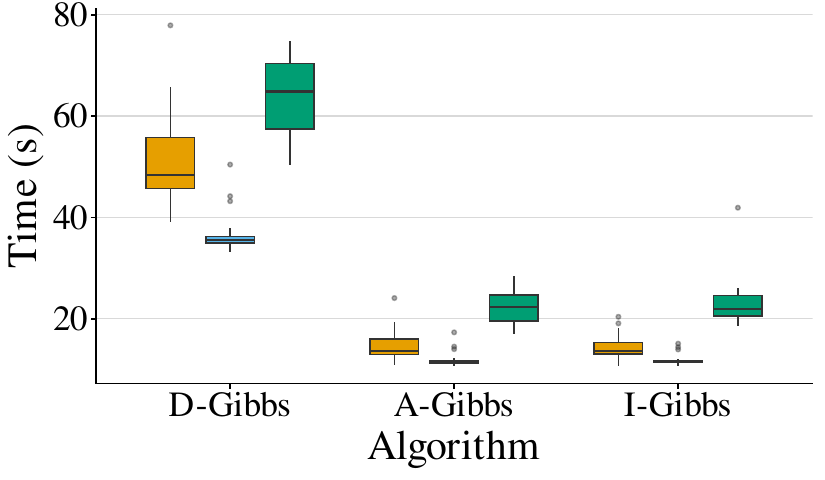}
        \caption{$(n,p)=(500,100)$}
    \end{subfigure}
    \begin{subfigure}[b]{0.32\textwidth}
        \centering
        \includegraphics[width=\textwidth, height = 0.12\textheight]{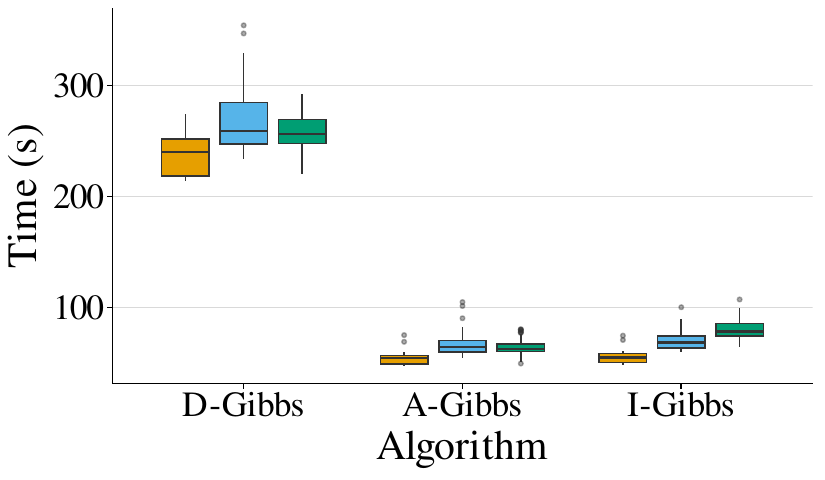}
        \caption{$(n,p)=(500,250)$}
    \end{subfigure}
    
    \caption{Comparison of average runtimes (in seconds) of the 3000 iterations of D-Gibbs, A-Gibbs and I-Gibbs samplers, averaged over 50 replications for 3 different models (line, grid and Erd\H{o}s-R\'enyi graph) under 6 different values of $(n,p)$.}
    \label{fig:time}
\end{figure}

As discussed previously in Lemma~\ref{lemma:4:compDirect} and~\ref{lemma:7:compaug}, the computational complexity of D-Gibbs is $\mathcal{O}(p^3)$, and that of A-Gibbs is $\mathcal{O}(\min\{np^{2},p^3\})$, which creates a computational bottleneck for D-Gibbs and provides A-Gibbs a significant advantage over it, especially when $p \gg n$. {\color{black} Furthermore, even when \(n>p\), for which D-Gibbs and A-Gibbs have the same asymptotic time complexity, Figure~\ref{fig:time} shows that D-Gibbs incurs a higher wall-clock runtime. This difference is primarily attributable to the dependence among the full conditional distributions of the entries of \(\Omega\), which necessitates sequential updates in D-Gibbs. In contrast, the proposed data augmentation scheme enables joint updates in A-Gibbs through optimized matrix operations implemented using efficient C++ linear algebra libraries.} We later introduced I-Gibbs to periodically induce the local behavior of D-Gibbs in the chain and avoid the computational burden simultaneously. Below, we provide a comparison of the average time needed to execute the 3 Gibbs samplers in line, grid and Erd\H{o}s-R\'enyi model for all the $6$ combinations of $(n,p)$. The average is taken over the $50$ datasets, and the full comparison is provided in Figure~\ref{fig:time}.   

Figure~\ref{fig:time} clearly shows the advantage of the A-Gibbs and I-Gibbs sampler over the D-Gibbs sampler in terms of their runtimes for all the settings, supporting the order of magnitude difference between the computational complexities of D-Gibbs and A-Gibbs. The runtimes of the A-Gibbs and the I-Gibbs sampler are quite similar for smaller $p$, and although for larger values of $p$, the runtime for I-Gibbs is slightly higher than A-Gibbs, it is still negligible. This is expected, as I-Gibbs replaces only 1\% of A-Gibbs iterations by D-Gibbs ones.

\subsection{Comparison of Area Under Receiver Operating Characteristic Curve (AUROC)} \label{App:B3:AUROC}

{\color{black}The Receiver Operating Characteristic (ROC) curve plots the true-positive rate against the false-positive rate across classification thresholds. The area under the ROC curve (AUROC) summarizes overall discriminatory performance, with values closer to 1 indicating better classification.

To compute the AUROC, we consider a sequence of $50$ hyperparameter values. For each of the 18 simulation settings (three data-generating models and six pairs of $(n,p)$ values), we evaluate AUROC over the $50$ hyperparameter values, and average them across 50 independently generated datasets. We restrict the ROC curve to the region FPR $ \leq 0.15$ and compute the corresponding partial AUROC. To facilitate comparisons on the unit scale, the resulting partial AUROC is normalized by dividing by $0.15$. The normalized AUROC values for GLASSO, CLIME, D-Gibbs, A-Gibbs, and I-Gibbs are reported in Table~\ref{tab:6:AUC}. GGL is intentionally omitted from this comparison because its intrinsic sparsity constraints enforce sparsity across the range of hyperparameter values and often prevent the FPR from reaching the 0.15 cutoff, and therefore preclude a directly comparable partial-AUROC calculation.

Across all $18$ settings the methods generally achieved high AUROC values. The sole exception occurred under the Erd\H{o}s–R\'enyi generating model with $(n,p)=(100,200)$, where every method performed worse than random guessing. Among the proposed samplers, \textbf{I‑Gibbs exhibited} \textbf{particularly strong performance} for the \textbf{grid} and \textbf{Erd\H{o}s–R\'enyi} models. Although CLIME generally yielded slightly higher AUROC values in the line-graph setting, the differences were negligible and typically appeared only beyond the third decimal place. A similar pattern holds for the few instances where GLASSO achieved a modest edge. Within our proposed family of Gibbs samplers, D-Gibbs generally exhibited the weakest performance, A-Gibbs yielded substantial improvements over D-Gibbs, and I-Gibbs provided a further improvement over A-Gibbs.}

\begin{table}[!ht]
\centering
{\setstretch{1.3}
\scriptsize
\caption{\textit{AUROC (Area Under the Receiver Operating Characteristic curve)} for the benchmark and proposed algorithms across three data-generating models, namely, Line Graph, Grid Graph, and Erdős–Rényi Graph ($5\%$ edge density), and six \((n,p)\) configurations.}
\label{tab:6:AUC}

\begin{tabular}{c c c c c c c}
\toprule
\textbf{Model} & $\textbf{(n,p)}$ & \textbf{GLASSO} & \textbf{CLIME} & \textbf{D-Gibbs} & \textbf{A-Gibbs} & \textbf{I-Gibbs} \\
\midrule
\multirow{6}{*}{Grid Graph} & $(100,20)$ & 0.554125 & 0.816922 & 0.780880 & 0.833244 & \textbf{0.842288} \\
 & $(100,50)$ & 0.755612 & 0.784037 & 0.747884 & 0.836661 & \textbf{0.851225} \\
 & $(100,100)$ & 0.819931 & 0.819204 & 0.697720 & 0.815764 & \textbf{0.828695} \\
 & $(100,200)$ &  \textbf{0.860535} & 0.809257 & 0.696910 & 0.812945 & 0.832086\\
 & $(500,100)$ &  0.950085 & 0.994948 & 0.990655 & 0.993990 & \textbf{0.996140} \\
 & $(500,250)$ &  0.983930 & 0.994093 & 0.984323 & 0.991919 & \textbf{0.994223} \\
\midrule
\multirow{6}{*}{Erd\H{o}s-R\'enyi} & $(100,20)$ & 0.936876 & 0.974328 & 0.815850 & 0.983847 & \textbf{0.986017} \\
& $(100,50)$ &  0.686814 & 0.749381 & 0.647868 & 0.758833 & \textbf{0.769337}\\
 & $(100,100)$ &  0.548982 & 0.512198 & 0.375121 & 0.541621 & \textbf{0.552572} \\
 & $(100,200)$ & \textbf{0.284322} & 0.217657 & 0.121556 & 0.273471 & 0.277320\\
 & $(500,100)$ & 0.841647 & 0.925679 & 0.878911 & 0.929387 & \textbf{0.940927}\\
 & $(500,250)$ &  0.529413 & 0.492594 & 0.313160 & 0.527501 & \textbf{0.535772} \\
 \midrule
\multirow{6}{*}{Line Graph} & $(100,20)$ & 0.811230 & 0.957063 & 0.947920 & \textbf{0.975754} & 0.974224 \\
 & $(100,50)$ & 0.928225 & \textbf{0.979901} & 0.950205 & 0.974808 & 0.976180 \\
 & $(100,100)$ & 0.960837 & \textbf{0.981858} & 0.948411 & 0.970938 & 0.973246 \\
 & $(100,200)$ & \textbf{0.954007} & 0.948304 & 0.915276 & 0.931987 & 0.939708\\
 & $(500,100)$ &  0.983510 & \textbf{0.999860} & 0.995035 & 0.990166 & 0.997524\\
 & $(500,250)$ &  0.996227 & \textbf{0.999957} & 0.997976 & 0.999267 & 0.999080 \\
\bottomrule
\end{tabular}
}
\end{table}

{\color{black}\section{Additional Details on Graphical Models Under Diagonal Dominance Constraint} \label{supp:sec:DD}

In this section, we extend the generalized Bayesian framework developed under the positivity constraint $\mathcal{S}^{*}(p)$ to the diagonally dominant class $\mathcal{S}^{**}(p)$. As discussed in Section~\ref{sec:3:model}, $\mathcal{S}^{**}(p)$ consists of symmetric matrices with positive diagonal entries, non-positive off-diagonal entries, and strict diagonal dominance, which guarantees positive definiteness. Diagonally dominant precision matrices arise naturally in several applications, including conditional autoregressive (CAR) models, multiresolution spatial graphical regression~\citep{chen2026multi}, and genetic regulatory networks~\citep{ZAVLANOS20111113}. These considerations motivate the development of scalable Bayesian procedures under the joint positivity and diagonal-dominance constraints. Building on the framework developed in the preceding sections, we develop a generalized Bayesian approach for this restricted class using the D-trace generalized likelihood introduced in Section~\ref{subsec:2.1:dtrace}. 

\subsection{Model Description} \label{supp:subsec:Model_DD}
In this section, we specify the prior distribution for $\Omega$ under the constraints defining $\mathcal{S}^{**}(p)$ and derive the corresponding joint generalized posterior distribution. As in Section~\ref{sec:3:model}, sparsity is induced through spike-and-slab prior components for the off-diagonal entries of $\Omega$, while flat prior components are used for the diagonal entries. Unlike the setting considered in Section~\ref{sec:3:model}, the diagonal-dominance constraints couple the entries of $\Omega$, and therefore the resulting prior is no longer componentwise independent. Specifically, we define
\[
f(\Omega)
\propto
\left\{
\prod_{i=1}^{p} f_{1}(\omega_{ii})
\prod_{1\le i<j\le p} f_{2}(\omega_{ij})
\right\}
I_{\mathcal{S}^{**}(p)}(\Omega),
\]
where $f_1$ denotes the flat prior component for the diagonal entries and $f_2$ denotes the spike-and-slab prior component for the off-diagonal entries. Specifically, $f_2$ consists of a point mass at zero and a truncated normal slab $\mathcal{N}(0,\tau^2;-\infty,0)$ with slab inclusion probability $q\in(0,1)$. The indicator $I_{\mathcal{S}^{**}(p)}(\Omega)$ further restricts the support of the joint prior to matrices satisfying the positivity and diagonal dominance constraints.
Under the generalized likelihood based on the D-trace loss, the priors described above, and the diagonal dominance and positivity constraints, the joint generalized posterior density of $\Omega$ is calculated as given below. 
\begin{equation}\label{eqn:S3:post:exact:DD}
\fontsize{11}{12}\selectfont
\pi_D^{DD}(\Omega\mid \Yb)
\propto
\exp\left\{
n\sum_{i=1}^{p}\omega_{ii}
-\frac{n}{2}\operatorname{tr}(\Omega^2 S)
\right\}
\prod_{\substack{i,j=1\\ i<j}}^{p}
\left[
(1-q)I_{\{0\}}(\omega_{ij})
+
q\frac{2}{\tau\sqrt{2\pi}}
\exp\left\{-\frac{\omega_{ij}^{2}}{2\tau^{2}}\right\}
\right]I_{ \mathcal{S}^{**}(p)}(\Omega). 
\end{equation}

 \begin{lemma}[Posterior propriety]
\label{lemma:8:post_prop_DD}
Let $S=n^{-1}\Yb^{T}\Yb$, and suppose that $s_{ii}>0$ for all $i=1,\ldots,p$. Then the joint generalized posterior distribution of $\Omega$ conditional on $\Yb$, given in \eqref{eqn:S3:post:exact:DD}, is proper; that is, its normalizing constant is finite.
\end{lemma}

The proof follows immediately from the proof of Lemma~\ref{lemma:2:post:proper} after replacing $\mathcal{S}^{*}(p)$ by its subset $\mathcal{S}^{**}(p)$, and is therefore omitted. Consequently, the generalized posterior distribution of $\Omega$ under the joint positivity and diagonal dominance constraints is proper and can be used for posterior sampling.

\subsection{Sampling from the Posterior} \label{supp:subsec:gibbs_DD}
Although direct sampling from the joint posterior distribution in~\eqref{eqn:S3:post:exact:DD} is intractable, the full conditional distributions admit closed-form truncated-normal representations. Consequently, following the strategy developed in Section~\ref{subsec:4.1:directgibbs}, we construct a componentwise direct Gibbs sampler (D-Gibbs) for posterior computation.
\begin{lemma}[Conditional Posterior Distribution] \label{lemma:8:post:direct_DD}
The conditional posterior distributions of the off-diagonal entries of $\Omega$, under the $M$-matrix and diagonal dominance constraints, given the remaining entries, are characterized by the following posterior densities. For $1 \leq i < j \leq p$,
\begin{align}\label{eqn:S4:fullcond:offdiag_DD}
    \omega_{ij}\mid \Yb, \Omega_{-(ij)} 
    \sim (1 - q_{ij}^{D}) \delta_0 
    + q_{ij}^D\, \mathcal{N}\!\left(-\frac{b_{ij}}{a_{ij}},\, \frac{1}{n a_{ij}};\, -\alpha_{ij}^{D}, 0 \right),
\end{align}
where
\[
q_{ij}^D = \frac{c_{ij}^{D}}{1 + c_{ij}^{D}}, \quad 
c_{ij}^{D} = \frac{2q}{(1 - q)\sqrt{n a_{ij} \tau^2}} \exp\!\left(\frac{n b_{ij}^2}{2 a_{ij}}\right)\left[\Phi\left(
b_{ij}\sqrt{\frac{n}{a_{ij}}}
\right)-\Phi\left(b_{ij}\sqrt{\frac{n}{a_{ij}}}-\alpha_{ij}^{D} \sqrt{na_{ij}}\right)\right],
\]
\[
a_{ij} = s_{ii} + s_{jj} + \frac{1}{n \tau^2}, \quad 
b_{ij} = \sum_{j' \neq j} \omega_{i j'} s_{j j'} + \sum_{i' \neq i} \omega_{i' j} s_{i i'}, \quad
\alpha_{ij}^{D} = \min(\sum_{j^{'} \neq j} \omega_{ij^{'}},\sum_{i^{'} \neq i} \omega_{i^{'}j}),
\]
and $s_{ij}$ denotes the $(i,j)$-th entry of $S = \frac{1}{n}\Yb^\top \Yb$. For any matrix $A$, $A_{-(ij)}$ denotes the vector of the entries in the upper triangle except the $(i,j)$-th entry.

The conditional posterior distributions of the diagonal entries of $\Omega$, given the remaining entries, under the specified constraints, are given by
\begin{align}\label{eqn:S5:fullcond:diag_DD}
    \omega_{ii}\mid \Yb, \Omega_{-(ii)} 
    \sim \mathcal{N}\!\left(\frac{1 - d_i}{s_{ii}},\, \frac{1}{n s_{ii}};\, \beta_{i}^{D}, \infty \right),
\end{align}
where $\displaystyle d_i = \sum_{i' \neq i} \omega_{i i'} s_{i i'}, \,\,\beta_{i}^{D} =  \sum_{j \neq i} |\omega_{ij}| = - \sum_{j \neq i} \omega_{ij} , \quad$ for $i = 1, 2, \dots, p.$
    
\end{lemma}

A schematic of the componentwise direct Gibbs sampler has been provided in Algorithm~\ref{algo:4:directgibbs_DD}.

\begin{algorithm}[!ht]
	\caption{Direct Gibbs Sampler} \label{algo:4:directgibbs_DD} 
    {\setstretch{1.22}
	\begin{algorithmic}[1]
	\Procedure{D-GIBBS($\Omega^{init},\Yb, T, q, \tau^2$)}{}
        \State $\Omega^{(0)}=\Omega^{init}, \, $
		\State $S = n^{-1}\Yb^{T} \Yb$
        \State $\alpha = ((\alpha_{ij}))_{i,j=1}^{p}$ with $\alpha_{ij} = \sum_{j' \neq j} \omega_{ij}$, $\beta = (\beta_{i})_{i=1}^{p}$ with $\beta_{i} = \sum_{i' \neq i} {|\omega_{ij}|} $
        \For{$iter=1,2,\dots,T$}
          \State $\Omega^{new} \gets \Omega^{(iter-1)}$
		\For {$i=1,2,\ldots,p-1$} \Comment{Updating off-diagonals of $\Omega$}
		   \For {$j=i+1,\ldots,p$}

\State $a \gets s_{ii}+s_{jj}+1/(n\tau^2), \quad
b \gets (\Omega^{new}_{.i})^T S_{.j}
+(\Omega^{new}_{.j})^T S_{.i}
-a\omega_{ij}^{new}$

\State $l \gets -\min(\alpha_{ij},\alpha_{ji}), \, H_{\Phi} \gets
\Phi\!\left(b\sqrt{\frac{n}{a}}\right)
-
\Phi\!\left(
b\sqrt{\frac{n}{a}}
+\alpha_{ij}\sqrt{na}
\right)$

\State $P(0)\gets 1, \quad
P(1)\gets
\frac{2q}{(1-q)\sqrt{na\tau^2}}
\exp\!\left(\frac{nb^2}{2a}\right)H_{\Phi}$
				\If {$P(1) = \infty$}
				\State $\omega^{new}_{ij} \gets \mathcal{N}\left(-\frac{b}{a},\frac{1}{na};l,0\right),\,\omega_{ji}^{new} \gets \omega_{ij}^{new}$ 
				\Else 
				\State $P \gets {P/\text{sum}(P)}$
				  \State $\omega_{ij}^{new} \sim P(0) \delta_0 + P(1) \mathcal{N}\left(-\frac{b}{a},\frac{1}{na};l,0\right),\,\omega_{ji}^{new} \gets \omega_{ij}^{new}$ 
                \EndIf
                \State $\Delta_{1} = \omega_{ij}^{new}-\Omega^{(iter-1)}_{ij} $
                \State $\beta_{i} \gets \beta_{i} - \Delta_{1},\, \beta_{j} \gets \beta_{j}-\Delta_{1}$ \Comment{Updating $\beta$}
                \State $(\alpha_{i.})_{-j} \gets (\alpha_{i.})_{-j} + \Delta_{1},\, (\alpha_{j.})_{-i} \gets (\alpha_{j.})_{-i} + \Delta_{1}$ \Comment{Updating $\alpha$}
		    \EndFor 
		\State $d \gets 1-(\Omega_{.i}^{new})^{T} S_{.i}+\omega_{ii}^{new}s_{ii} $ \Comment{Updating diagonals of $\Omega$}
		\State $\omega_{ii}^{new} \gets \mathcal{N}(\frac{d}{s_{ii}},\frac{1}{ns_{ii}};\beta_{i},\infty)$ 
        \State $\Delta_{2} \gets \omega_{ii}^{new} - \Omega^{(iter-1)}_{ii}, \alpha_{i.} \gets \alpha_{i.} + \Delta_{2} $ \Comment{Updating $\alpha$}
		\EndFor
		\State Repeat Steps 22 - 24 for $i = p$
        \State $\Omega^{(iter)} \gets \Omega^{new}$
         \EndFor
		\State \textbf{return} $\Omega^{(1)},\Omega^{(2)}, \dots,\Omega^{(T)}$
	\EndProcedure
	\end{algorithmic}
    }
\end{algorithm}
The updates required to maintain the diagonal-dominance constraints can be implemented incrementally after each entry of $\Omega$ is sampled. In particular, the updates of $\alpha$ and $\beta$ involve only scalar corrections to previously computed quantities and therefore do not alter the leading-order computational cost. Consequently, a complete Gibbs sweep still requires $O(p^3)$ operations. See Steps~20 and~24 of Algorithm~\ref{algo:4:directgibbs_DD} for implementation details. 

We next develop an augmented Gibbs sampler (A-Gibbs) for the present setting by introducing the auxiliary matrix $R$ as in Section~\ref{subsec:4.2:aug:gibbs}. Recall that, conditional on $\Yb$ and $\Omega$,
\[
R\mid \Yb,\Omega
\sim
\mathcal{MN}_{p\times p}
\bigl(
\Omega(kI_p-nS),\,
I_p,\,
kI_p-nS
\bigr).
\]
When $p>n$, samples from this matrix-normal distribution can be generated efficiently by exploiting its Gram structure through the fast sampling scheme of Lemma~\ref{lemma:6:struct:gauss}. Under this augmentation, the conditional distributions retain tractable forms analogous to those derived in Section~\ref{subsec:4.2:aug:gibbs}, with the diagonal-dominance constraints appearing through the truncation limits. 
\begin{lemma}[Conditional posterior distribution Under Augmentation]
\label{lemma:9:augcond_DD}
Under the augmented formulation in \eqref{eqn:7:distR}, the conditional posterior densities of the entries of $\Omega$, conditioned on $R$, $\Yb$, and the remaining entries of $\Omega$ are as follows. For $1\le i<j\le p$,
\begin{align}
\label{eqn:S7:augcond_DD}
    \omega_{ij}\mid R,\Yb,\Omega_{-(ij)}
    &\sim (1-q_{ij}^{*D})\delta_{0}
    +q_{ij}^{*D}\mathcal{N}\left(
    \frac{R_{ij}+R_{ji}}{k^{*}},
    \frac{1}{k^{*}};
    -\alpha_{ij}^{D},0
    \right), \nonumber\\
    \omega_{ii}\mid R,\Yb,\Omega_{-(ii)}
    &\sim \mathcal{N}\left(
    \frac{R_{ii}+n}{k},
    \frac{1}{k};
    \beta_{i}^{D}, \infty
    \right),
\end{align}
where
\[
\alpha_{ij}^{D}
=
\min\left(
\sum_{j^{'} \neq j} \omega_{ij^{'}},
\sum_{i^{'} \neq i} \omega_{i^{'}j}
\right),\,\,
\beta_{i}^{D}
=
-\sum_{j \neq i} \omega_{ij},\,\,
k^{*}=2k+\frac{1}{\tau^{2}},\,\,
q_{ij}^{*D}=\frac{c_{ij}^{D}}{1+c_{ij}^{D}},
\]
and
\[
c_{ij}^{D}
=
\frac{2q}{1-q}\frac{1}{\tau\sqrt{k^{*}}}
\exp\left\{
\frac{(R_{ij}+R_{ji})^{2}}{2k^{*}}
\right\}
\left[
\Phi\left(
-\frac{R_{ij}+R_{ji}}{\sqrt{k^{*}}}
\right)
-
\Phi\left(
-\frac{R_{ij}+R_{ji}}{\sqrt{k^{*}}}
-\alpha_{ij}^{D}\sqrt{k^{*}}
\right)
\right].
\]
Unlike the positivity-constrained formulation considered in Section~\ref{subsec:4.2:aug:gibbs}, conditioning on $R$ does not render the entries of $\Omega$ mutually independent, since the diagonal-dominance constraints induce dependence through the truncation bounds. 
\end{lemma}

Lemma~\ref{lemma:8:post:direct_DD} and Lemma~\ref{lemma:9:augcond_DD} can be established along similar lines as the proofs of Lemma~\ref{lemma:3:post:comp} and Lemma~\ref{lemma:5:augcond}, respectively; see Supplemental Sections~\ref{app:A2:pf:l3} and~\ref{app:A5:pf:l6}. Their proofs are therefore omitted. A schematic of the resulting A-Gibbs sampler is provided in Algorithm~\ref{algo:5:auggibbs_DD} below.

\begin{algorithm}[!ht]
	\caption{Augmented Gibbs Sampler} \label{algo:5:auggibbs_DD} 
    {\setstretch{1.22}
	\begin{algorithmic}[1]
	\Procedure{A-GIBBS($\Omega^{init},\Yb,T,\varepsilon,q,\tau^2$)}{}
        \State $\Omega^{(0)} = \Omega^{init}$
		\State $S = n^{-1}\Yb^{T} \Yb$, $\lambda=$ Maximum eigenvalue of $nS$ 
        \State $\alpha = ((\alpha_{ij}))_{i,j=1}^{p}$ with $\alpha_{ij} = \sum_{j' \neq j} \omega_{ij}$, $\beta = (\beta_{i})_{i=1}^{p}$ with $\beta_{i} = \sum_{i' \neq i} {|\omega_{ij}|} $
        \State $k \gets \varepsilon \lambda,\,k^{*} \gets 2k+\tau^{-2}$
        \For{iter $=1,2,\dots,T$}
        \If {$p> n$}  \Comment{Sampling Augmented variable $R$}
        \State $R \gets \mathcal{MN}_{p \times p}\bigl(\Omega^{(iter-1)}(kI_p-nS),\,I_p,\,kI_p-nS\bigr)$ using Lemma~\ref{lemma:6:struct:gauss}
        \Else 
        \State $R \gets \mathcal{MN}_{p \times p}\bigl(\Omega^{(iter-1)}(kI_p-nS),\,I_p,\,kI_p-nS\bigr)$ directly
        \EndIf
        \State $\Omega^{new} \gets \Omega^{(iter-1)}$
		\For {$i=1,2,\ldots,p-1$} \Comment{Updating off-diagonals of $\Omega$}
		   \For {$j=i+1,\ldots,p$}
		      \State $l \gets -\min(\alpha_{ij},\alpha_{ji}), \,H_{\Phi} \gets 
\Phi\!\left(-\frac{R_{ij}+R_{ji}}{\sqrt{k^{*}}}\right)
-
\Phi\!\left(
-\frac{R_{ij}+R_{ji}}{\sqrt{k^{*}}}
+\alpha_{ij}\sqrt{k^{*}}
\right)$

\State $P(0) \gets 1,\quad 
P(1) \gets 
2\sqrt{\frac{1}{\tau^{2}k^{*}}}
\frac{q}{1-q}
\exp\!\left[\frac{(R_{ij}+R_{ji})^2}{2k^{*}}\right]H_{\Phi}$
                \If {$P(1) = \infty$}
				\State $\omega_{ij}^{new} \gets \mathcal{N}\left(\frac{R_{ij}+R_{ji}}{k^{*}},\frac{1}{k^{*}};l,0\right),\,\omega_{ji}^{new} \gets \omega_{ij}^{new}$
				\Else 
				\State $P \gets {P/\text{sum}(P)}$ \State $\omega_{ij}^{new} \sim P(0) \delta_0 + P(1) \mathcal{N}\left(\frac{R_{ij}+R_{ji}}{k^{*}},\frac{1}{k^{*}};l,0\right),\,\omega_{ji}^{new} \gets \omega_{ij}^{new}$
				\EndIf
                \State $\Delta_{1} = \omega_{ij}^{new}-\Omega^{(iter-1)}_{ij} $
                \State $\beta_{i} \gets \beta_{i} - \Delta_{1},\, \beta_{j} \gets \beta_{j}-\Delta_{1}$ \Comment{Updating $\beta$}
                \State $(\alpha_{i.})_{-j} \gets (\alpha_{i.})_{-j} + \Delta_{1},\, (\alpha_{j.})_{-i} \gets (\alpha_{j.})_{-i} + \Delta_{1}$ \Comment{Updating $\alpha$}
		    \EndFor 
		\State $\omega_{ii}^{new} \gets \mathcal{N}(\frac{R_{ii}+n}{k},\frac{1}{k};\beta_{i},\infty)$  \Comment{Updating diagonals of $\Omega$}
        \State $\Delta_{2} \gets \omega_{ii}^{new} - \Omega^{(iter-1)}_{ii}, \alpha_{i.} \gets \alpha_{i.} + \Delta_{2} $ \Comment{Updating $\alpha$}
		\EndFor
		\State Repeat Steps 27-28 for $i = p$
		\State $\Omega^{(iter)} \gets \Omega^{new}$ 
        \EndFor
        \State \textbf{return} $\Omega^{(1)},\Omega^{(2)},\dots,\Omega^{(T)}$ 
	\EndProcedure	
	\end{algorithmic}
    }
\end{algorithm}

The conditional-independence structure exploited in Section~\ref{subsec:4.2:aug:gibbs} is lost under diagonal dominance. As a result, the computational complexity of A-Gibbs increases from $O(np^2)$ to $O(p^3)$ when $n\ll p$. This additional cost is the price of enforcing diagonal dominance and thereby guaranteeing positive-definite precision matrix draws. Nevertheless, in our numerical experiments (Section \ref{supp:subsec:sim_DD}), the D-Gibbs sampler exhibits occasional instability over a small number of iterations, particularly for larger $n$, whereas A-Gibbs remains comparatively stable across the $(n,p)$ configurations considered. Moreover, as discussed in detail in Section~\ref{subsec:4.3:interweaving}, A-Gibbs facilitates broader exploration through the augmented parameter space, while D-Gibbs performs more localized updates in the original parameter space. Motivated by these complementary behaviors, similar to Section \ref{subsec:4.3:interweaving}, we propose an Interweaving Gibbs sampler that combines both mechanisms by inserting one D-Gibbs update within a sequence of A-Gibbs updates. A schematic of the resulting interweaving Gibbs sampler is provided in Algorithm \ref{algo:6:interweaving_DD}.

\begin{algorithm}[!ht]
{\setstretch{1.22}
	\caption{Interweaving Gibbs Sampler} \label{algo:6:interweaving_DD} 
	\begin{algorithmic}[1]
	\Procedure{I-GIBBS($\Omega^{init},\Yb,T,\varepsilon,q,\tau^2,N$)}{}
        \State $\Omega^{(0)}=\Omega^{init}$
		\State $S = n^{-1}\Yb^{T} \Yb$, $\lambda=$ Maximum eigenvalue of $nS$ 
        \State $ \alpha = ((\alpha_{ij}))_{i,j=1}^{p}$ with $\alpha_{ij} = \displaystyle \sum_{j' \neq j} \omega_{ij}$, $\beta = (\beta_{i})_{i=1}^{p}$ with $\beta_{i} = \displaystyle \sum_{i' \neq i} {|\omega_{ij}|} $
        \For{$iter =1,2,\dots,T$}
          \If{\(\mathrm{iter}\equiv 0 \pmod N\)}
               \State Perform Steps 6-27 of \textsc{D-GIBBS} in Algorithm~\ref{algo:4:directgibbs_DD} with $\Omega^{init}=\Omega^{(iter-1)}$
            \Else
               \State Perform Steps 7-31 of \textsc{A-GIBBS} in Algorithm~\ref{algo:5:auggibbs_DD} with $\Omega^{init}=\Omega^{(iter-1)}$
            \EndIf
            \State $\Omega^{(iter)} \gets$ sampled matrix in Steps 6-10
        \EndFor
        \State \textbf{return} $\Omega^{(1)},\Omega^{(2)},\dots,\Omega^{(T)}$ 
	\EndProcedure	
	\end{algorithmic}
    }
\end{algorithm}

Once the post-burn-in posterior samples of $\Omega$ are obtained from any of the three proposed algorithms, the graph structure can be estimated using the majority-voting approach and final precision matrix estimate is then obtained by averaging over the post burn-in samples as described in Section~\ref{subsec:4.5:final_estimation}.

\subsection{Numerical Experiments} \label{supp:subsec:sim_DD}
We evaluate the proposed samplers for graph recovery and precision matrix estimation under diagonal dominance, comparing them with GLASSO, CLIME, and DDGL, the diagonally dominant analogue of GGL. Since the true precision matrix corresponding to the line-graph generating model in Section~\ref{sec:5:simulation} is diagonally dominant, we use the same line-graph setting and synthetic data-generation scheme considered earlier. For each fixed $(n,p)$ configuration, 50 independent datasets are generated with $6n/5$ observations, of which $n$ are used for model fitting and the remaining $n/5$ serve as test data for tuning. For the proposed samplers, the slab inclusion probability $q$ is selected by grid search over $\{1/p,1/p^{1.5},1/p^2\}$ using the held-out data, as described in Section~\ref{subsec:4.4:hyper}, while the regularization parameters of the benchmark methods are selected by cross-validation over the ranges recommended by their respective authors. The resulting MCC, F-score, and relative estimation errors are reported in Tables~\ref{tab:7:MCC_DD},~\ref{tab:8:Fscore_DD}, and~\ref{tab:9:est_error_DD}, respectively.

\begin{table}[!ht]
\centering
{\setstretch{1.3}
\footnotesize
\caption{\textit{Matthews correlation coefficient (MCC)} for the benchmark and proposed algorithms across six \((n,p)\) configurations, where \(n\) and \(p\) denote the sample size and number of variables, respectively, under the Line Graph generating model under \textbf{diagonal dominance constraint} on the precision matrix.}
\label{tab:7:MCC_DD}
\begin{tabular}{c c c c c c c}
\toprule
$\textbf{(n,p)}$ & \textbf{GLASSO} & \textbf{CLIME} & \textbf{DDGL} & \textbf{D-Gibbs} & \textbf{A-Gibbs} & \textbf{I-Gibbs} \\
\midrule
$(100,20)$ & $0.339362$ & $0.162940$ & $0.575641$ & $0.936973$ & $0.925525$ & $\textbf{0.938925}$ \\
$(100,50)$ & $0.344382$ & $0.366485$ & $0.486751$ & $\textbf{0.921028}$ & $0.916477$ & $0.919972$ \\
$(100,100)$ & $0.338153$ & $0.469339$ & $0.420468$ & $0.896818$ & $0.883941$ & $\textbf{0.896887}$ \\
$(100,200)$ & $0.340614$ & $0.545414$ & $0.357547$ & $0.846959$ & $0.811773$ & $\textbf{0.852325}$ \\
$(500,100)$ & $0.324436$ & $0.442767$ & $0.390768$ & $0.909015$ & $0.951251$ & $\textbf{0.956816}$\\
$(500,250)$ & $0.328980$ & $0.472171$ & $0.309920$ & $0.950397$ & $0.965262$ & $\textbf{0.965424}$\\
\bottomrule
\end{tabular}
}
\end{table}

\begin{table}[!ht]
\centering
{\setstretch{1.3}
\footnotesize
\caption{\textit{$F$-scores} for the benchmark and proposed algorithms across six \((n,p)\) configurations, where \(n\) and \(p\) denote the sample size and number of variables, respectively, under the Line Graph generating model under \textbf{diagonal dominance constraint} on the precision matrix.}
\label{tab:8:Fscore_DD}
\begin{tabular}{c c c c c c c}
\toprule
 $\textbf{(n,p)}$ & \textbf{GLASSO} & \textbf{CLIME} & \textbf{DDGL} & \textbf{D-Gibbs} & \textbf{A-Gibbs} & \textbf{I-Gibbs} \\
\midrule
 $(100,20)$ & $0.346516$ & $0.223613$ & $0.569840$ & $0.942625$ & $0.931548$ & $\textbf{0.944377}$ \\
 $(100,50)$ & $0.269845$ & $0.290592$ & $0.422301$ & $0.923891$ & $0.919400$ & $\textbf{0.924868}$ \\
 $(100,100)$ & $0.235832$ & $0.384320$ & $0.324350$ & $0.898747$ & $0.885914$ & $\textbf{0.898773}$ \\
 $(100,200)$ & $0.227745$ & $0.482143$ & $0.242386$ & $0.846669$ & $0.813571$ & $\textbf{0.846946}$\\
 $(500,100)$ & $0.222446$ & $0.350902$ & $0.290109$ & $0.909273$ & $0.952057$ & $\textbf{0.956747}$\\ 
 $(500,250)$ & $0.208240$ & $0.372885$ & $0.187231$ & $0.949986$ & $0.964967$ & $\textbf{0.965110}$\\
\bottomrule
\end{tabular}
}
\end{table}

\begin{table}[!ht]
\centering
{\setstretch{1.3}
\footnotesize
\caption{\textit{Relative Estimation Errors} (in Frobenius norm) for the benchmark and proposed algorithms across six \((n,p)\) configurations, where \(n\) and \(p\) denote the sample size and number of variables, respectively, under the Line Graph generating model under \textbf{diagonal dominance constraint} on the precision matrix.}
\label{tab:9:est_error_DD}

\begin{tabular}{c c c c c c c}
\toprule
$\textbf{(n,p)}$ & \textbf{GLASSO} & \textbf{CLIME} & \textbf{DDGL} & \textbf{D-Gibbs} & \textbf{A-Gibbs} & \textbf{I-Gibbs} \\
\midrule
 $(100,20)$ & $0.402249$ & $0.580012$ & $0.247147$ & $0.165747$ & $0.161152$ & $\textbf{0.160990}$\\
 $(100,50)$ & $0.537159$ & $0.435991$ & $0.297830$ & $0.180423$ & $\textbf{0.172346}$ & $0.174056$ \\
 $(100,100)$ & $0.577636$ & $0.416646$ & $0.341433$ & $0.201587$ & $0.196951$ & $\textbf{0.193663}$ \\
 $(100,200)$ & $0.564507$ & $0.395472$ & $0.370931$ & $0.229232$ & $0.237152$ & $\textbf{0.228886}$\\
 $(500,100)$ & $0.389555$ & $0.130876$ & $0.119532$ & $0.368408$  & $0.074493$ & $\textbf{0.074262}$\\
 $(500,250)$ & $0.388881$ & $0.160640$ & $0.136523$ & $0.228786$ & $\textbf{0.079839}$ & $0.080578$\\
\bottomrule
\end{tabular}
}
\end{table}

It is worth noting that, the proposed Gibbs samplers consistently achieve better sparsity recovery and precision matrix estimation than the benchmark methods, with substantial improvements in MCC, F-score, and relative estimation error. Among the three proposed samplers, I-Gibbs generally performs best, outperforming both D-Gibbs and A-Gibbs across the considered settings. As discussed in Section~\ref{supp:subsec:gibbs_DD}, D-Gibbs exhibits instability in a few simulation runs, particularly for $n=500$, which is reflected in the relative estimation errors reported in Table~\ref{tab:9:est_error_DD}. By alternating between D-Gibbs and A-Gibbs updates, the interweaving I-Gibbs sampler mitigates this instability while retaining the complementary local and global exploration of the two samplers, leading to improved overall performance. Moreover, the diagonally-dominant formulations consistently outperform their positivity-constrained counterparts across nearly all settings and evaluation metrics. This behavior is expected, since the data-generating precision matrix satisfies diagonal dominance, allowing the additional structural constraint to improve estimation efficiency and graph recovery accuracy.}

\section{Additional Details on Financial Data Analysis in Section~\ref{subsec:5.2:DataAnalysis}} \label{app:D:DA_uncertainty}

{\color{black} In this section, we report the marginal inclusion probabilities for all possible edges in the graphs estimated using our proposed Gibbs samplers. We partition the interval [0,1] into four equal-width intervals: [0,0.25), [0.25,0.5), [0.5,0.75), and [0.75,1]. The edges are further classified as within-sector or between-sector. After discarding the first 2000 iterations as burn-in, we compute the marginal inclusion probability of each edge as the proportion of the remaining 1000 posterior samples in which that edge is included. Table~\ref{tab:7:uncertainty} reports the resulting inclusion probabilities for each edge class.

D-Gibbs shows a clear trade-off between sensitivity to \textit{within-sector} structure and control of spurious \textit{between-sector} connections. It assigns more \textit{within-sector} edges to the high-probability intervals [0.5,0.75) and [0.75,1] than A-Gibbs and I-Gibbs, indicating stronger recovery of within-sector connectivity. This gain, however, comes at the cost of a substantially larger number of high-probability \textit{between-sector} edges. In contrast, A-Gibbs and I-Gibbs are more selective, retaining fewer between-sector connections while preserving the dominant within-sector structure. This sharper separation between within-sector and between-sector connectivity is reflected in their markedly higher modularity compared with D-Gibbs.
}

\begin{sidewaystable}[hbtp]
{\setstretch{1.3}
\centering
\small
\caption{Table of \textit{number of edges} belonging to different ranges of 
\textit{Proportion of within-sector edges}. The proportion of \textit{within-sector} 
and \textit{between-sector} edges out of the $1000$ posterior samples (after 
discarding the first $2000$ samples as burn-in) has been divided into 4 separate 
classes, and the number of edges in the estimated precision matrix belonging to 
each of the classes has been reported for all 3 proposed Gibbs samplers.}
\label{tab:7:uncertainty}

\begin{tabular}{c cc cc cc}
\toprule
\textbf{Proportion Range} 
& \multicolumn{2}{c}{\textbf{D-Gibbs}} 
& \multicolumn{2}{c}{\textbf{A-Gibbs}} 
& \multicolumn{2}{c}{\textbf{I-Gibbs}} \\
\cline{2-7}
& \textbf{Within-sector} & \textbf{Between-sector}
& \textbf{Within-sector} & \textbf{Between-sector}
& \textbf{Within-sector} & \textbf{Between-sector} \\
\midrule
0 -- 0.25 
& 8031 & 31073 
& 8353 & 32107 
& 8155 & 32075 \\
0.25 -- 0.5 
& 64 & 285 
& 4 & 2 
& 10 & 9 \\
0.5 -- 0.75 
& 43 & 100 
& 2 & 2 
& 9 & 4 \\
0.75 -- 1 
& 434 & 725 
& 213 & 72 
& 398 & 95 \\
\bottomrule
\end{tabular}

}
\end{sidewaystable}

\newpage

\end{document}